\documentclass[aps,pra,onecolumn,letterpaper,groupedaddress,10pt,superscriptaddress]{revtex4-2}
\usepackage{amssymb,amsthm,amsmath,amsfonts}
\usepackage{graphicx,ulem,enumerate,mathptmx,bbm}
\usepackage[extra]{tipa}
\usepackage[pdftex,dvipsnames,usenames]{xcolor}
\usepackage{tikz-cd}
\usepackage[colorlinks=true,urlcolor=blue,citecolor=blue,linkcolor=blue]{hyperref}
\usepackage[capitalize]{cleveref}

\begin{document}

\cleardoublepage

\title{Numerical approaches to entangling dynamics from variational principles}

\author{Christian Offen}
    \affiliation{School of Mathematics, University of Birmingham, Birmingham, B15 2TT, UK}
    
\author{Boris Wembe}
    \affiliation{Department of Mathematics, Paderborn University, Warburger Stra\ss{}e 100, D-33098 Paderborn, Germany}

\author{Laura Ares}
    \affiliation{Theoretical Quantum Science, Institute for Photonic Quantum Systems (PhoQS), Paderborn University, Warburger Stra\ss{}e 100, 33098 Paderborn, Germany}

\author{Jan Sperling}
    \affiliation{Theoretical Quantum Science, Institute for Photonic Quantum Systems (PhoQS), Paderborn University, Warburger Stra\ss{}e 100, 33098 Paderborn, Germany}

\author{Sina Ober-Bl\"obaum}
    \affiliation{Department of Mathematics, Paderborn University, Warburger Stra\ss{}e 100, D-33098 Paderborn, Germany}

\begin{abstract}

    In this work, we address the numerical identification of entanglement in dynamical scenarios.
    To this end, we consider different programs based on the restriction of the evolution to the set of separable (i.e., non-entangled) states, together with the discretization of the space of variables for numerical computations.
    As a first approach, we apply linear splitting methods to the restricted, continuous equations of motion derived from variational principles. 
    We utilize an exchange interaction Hamiltonian to confirm that the numerical and analytical solutions coincide in the limit of small time steps.
    The application to different Hamiltonians shows the wide applicability of the method to detect dynamical entanglement.
    To avoid the derivation of analytical solutions for complex dynamics, we consider variational, numerical integration schemes, introducing a variational discretization for Lagrangians linear in velocities. 
    Here, we examine and compare two approaches:
    one in which the system is discretized before the restriction is applied,
    and another in which the restriction precedes the discretization.
    We find that the ``first-discretize-then-restrict'' method becomes numerically unstable, already for the example of an exchange-interaction Hamiltonian, which can be an important consideration for the numerical analysis of constrained quantum dynamics.
    Thereby, broadly applicable numerical tools, including their limitations, for studying entanglement over time are established for assessing the entangling power of processes that are used in quantum information theory.

\end{abstract}

\maketitle
\section{Introduction}

    Determining whether a quantum state at a fixed time is separable or inseparable is known to be an NP-hard problem \cite{I07}. 
    In the special case of two and three-dimensional bipartite systems, necessary and sufficient criteria exist, allowing for efficient identification of entangled states \cite{P96}. 
    In contrast, high-dimensional and multipartite systems introduce additional intricacies, for example, because of the existence of multiple partitions, making criteria for entanglement certification significantly more challenging to derive \cite{H09}.
    To address these challenges in an experimentally feasible manner, analytical criteria in the form of entanglement witnesses have been developed \cite{GT09}.
    One well-established framework to derive optimal witnesses is through the separability eigenvalue equations \cite{SV13}.
    However, the sophistication of the separability problem reduces the availability of known exact solutions to limited scenarios.
    For this reason, numerical approaches for solving the separability eigenvalue equations have been investigated to help with the detection of multipartite entanglement in complex systems \cite{GVS18}.

    A limitation of the aforementioned approaches is that they apply to static scenarios only, probing entanglement at fixed times.
    Also, one ought to expect a speedup in terms of physical time, not only complexity, when a quantum processor has access to entangling dynamics, compared to a non-entangling evolution \cite{YS24}.
    Therefore, to certify entanglement in dynamical scenarios, several approaches have been developed over the last years; see, e.g., Refs. \cite{PhysRevLett.99.120501,PhysRevLett.101.170502,doi:10.1142/S1230161209000153,PhysRevA.65.012101,CLMOW14}.
    It is not uncommon to find relations analyzing the difference between the entanglement of initial and final states, resulting in an input-output-based quantum channel characterization with respect to its entangling power \cite{CLMOW14,PhysRevA.86.050302,PhysRevA.88.042335,PhysRevResearch.4.013200,pinske2025entanglingpowernonentanglingchannels}. 
    However, any entanglement at intermediate times is thereby not detected, especially when the initial and final states are separable \cite{SW20}.
    For this reason, variational principles were utilized to witness time-dependent entanglement \cite{SW17}, which recently have been generalized to open-system dynamics \cite{PA24,AP24}.
    Already in the closed-system approach, however, it becomes increasingly challenging to find exact solutions.
    Thus, there is a need for numerical methods, applicable to multipartite and high-dimensional systems, to benchmark the entangling part of the evolution of a composite quantum system.
    Further, numerical investigations, as carried out in the following, can even avoid the derivation of sophisticated equations of motion, dubbed separability Schr\"odinger equations (SSE), for the entangling evolution.
    
    In this work, we develop numerical methods for the identification of time-dependent entanglement by restricting the dynamics to be in a separable state for all times.
    First, we apply linear splitting methods to the restricted, continuous equations of motion derived from variational principles in Ref. \cite{SW17}.
    We gauge the accuracy of the method by comparing the numerical and analytical solutions for a proof-of-concept Hamiltonian and demonstrate the general applicability by detecting entanglement for different Hamiltonians and various system sizes.
    While numerical splitting methods are a natural choice for integrating the SSE, cf. Eq. \eqref{eq:SSE}, these require analytically deriving the continuous equations first. 
    To bypass this step, we propose a variational discretization approach that directly yields integration schemes for the restricted dynamics \cite{MarsdenWest2001,EllisonEtAl2018,OberBloebaum2015}.
    We consider two scenarios in this context: restrict–discretize and discretize–restrict. 
    In the first approach, the restriction is applied before the discretization whereas, in the second one, the order is reversed. 
    We compare the two methods and show that the discretize–restrict approach becomes numerically unstable when applied to the exchange-interaction Hamiltonian, highlighting a critical issue for the numerical treatment of constrained systems in numerical analysis.

\section{Separable and inseparable quantum dynamics}

    In this section, we very briefly review the method of SSE \cite{SW17,SW20} for characterizing the entangling capabilities of a Hamiltonian process.
    We consider $N$ complex Hilbert spaces $\mathcal{H}_1$, \ldots, $\mathcal{H}_N$ of dimension $d$, their tensor product $\mathcal{H}=\bigotimes_{j=1}^N \mathcal{H}_j$, and their Cartesian product $\mathcal{H}^\times =\prod_{j=1}^N \mathcal{H}_j$.
    An $N$-partite pure state  $|\psi\rangle\in \mathcal{H}$ is considered to be (partially or fully) entangled if it is not separable, i.e., if it cannot be expressed as a tensor product of pure states on each subsystem, $|\psi\rangle\notin \bigotimes^N_{j=1}|a_j\rangle$ with $|a_j\rangle\in \mathcal{H}_j$.

    To describe the evolution of a system, we can obtain its equations of motion from principle of stationary action \cite{GR96}. 
    In this way, we can recover the Schr\"odinger equation (SE),
    \begin{equation}\label{eq:SE}
         |\dot\psi\rangle = -\frac{i}{\hbar} \hat H |\psi \rangle,
    \end{equation}
    by minimizing the action $S=\int_0^T dt\,L$,  given the Lagrangian
	\begin{align}
		L=\frac{i\hbar}{2}\langle \psi|\dot\psi\rangle-\frac{i\hbar}{2}\langle\dot\psi|\psi\rangle-\langle \psi|\hat H|\psi\rangle,
	\end{align} 
    for a Hermitian operator $\hat H \colon \mathcal{H} \to \mathcal{H}$, 
    the Hamiltonian.
    This framework enables the systematic incorporation of constraints into the evolution of the system, for instance, by restricting trajectories to the preferred set of classical states \cite{SW20}.
    In particular, considering entanglement, we can restrict the state to evolve via product states, i.e., $|\psi (t)\rangle\in\mathcal{H^\times}$, not allowing for the presence of entanglement at any time $t$.
    To this end, we consider $|\psi(t)\rangle=\prod_{j=1}^N |a_j(t)\rangle$ and obtain the Euler-Lagrange equations 
    \begin{equation}\label{eq:SepEL}
		0=\frac{d}{dt}\frac{\partial L}{\partial \langle \dot a_j|}-\frac{\partial L}{\partial \langle a_j|},
	\end{equation}
    for each subsystem $j\in\{1,\ldots,N\}$.
    This program leads to the aformentioned SSE \cite{SW17}
    \begin{equation}\label{eq:SSE}
        \begin{split}
            |\dot a_1\rangle &= -\frac{i}{\hbar} \hat H_{a_2,\ldots,a_N} |a_1\rangle\\
            &\vdots\\
            |\dot a_k\rangle &= -\frac{i}{\hbar} \hat H_{a_1,\ldots,a_{k-1},a_{k+1},\ldots,a_N} |a_k\rangle\\
            &\vdots\\
            |\dot a_N\rangle &= -\frac{i}{\hbar} \hat H_{a_1,\ldots,a_{N-1}} |a_N\rangle,
        \end{split}
    \end{equation}
    where $\hat H_{a_1,\ldots,a_{k-1},a_{k+1},\ldots,a_N} \colon \mathcal{H}_k \to  \mathcal{H}_k$ denotes the $k$th partially reduced Hamiltonian 
    
        \begin{equation}
            \label{eq:partiallyredicedH}
    		\hat {H}_{a_1,\ldots,a_{k-1},a_{k+1},\ldots,a_N} =\frac{
    			\left(
    				\langle a_1,\ldots,a_{k-1}|
    				\otimes\hat 1_{\mathcal H_k}\otimes
    				\langle a_{k+1},\ldots,a_N|
    			\right)
    			\hat{H}
    			\left(
    				|a_1,\ldots,a_{k-1}\rangle
    				\otimes\hat 1_{\mathcal H_k}\otimes
    				| a_{k+1},\ldots,a_N\rangle
    			\right)
    		}{
    			\langle a_1\ldots a_{k-1}| a_1\ldots a_{k-1}\rangle
    			\langle a_{k+1}\ldots a_n| a_{k+1}\ldots a_n\rangle
    		},
    	\end{equation}

    where $\hat 1_{\mathcal H_k}$ is the identity for the subsystem $K$.
    (Note that a physically irrelevant, global-phase contribution, proportional to $\langle a_j|\dot a_j\rangle$, has been omitted in the SEE as given above, as formulated in the supplemental material of Ref. \cite{SW17}.)
    
    With this approach, one can benchmark dynamical entanglement in the actual process by comparing the separability-restricted and unrestricted trajectories;
    i.e., it represents a witness for time-dependent entanglement.
    It has been applied to derive separable speed limits \cite{YS24}, which are generally beaten by entangling dynamics, and has been extended to open systems \cite{PA24,AP24}.
    Nevertheless, the SSE being a set of nonlinear, coupled equations, obtaining analytical solutions for the SSEs \eqref{eq:SSE} becomes a challenge in most cases, emphasizing the need for robust numerical solvers.

\section{Numerical solution based on splitting method}
\label{sec:NumericsSplitting}

    To compute numerical solutions of the SE, we chose orthonormal bases of the Hilbert spaces, $\mathcal{H}_k=\mathrm{span}\{|0\rangle,\dots,|d-1\rangle\}$ for $k\in\{1,\ldots,N\}$, and consider the induced basis on the tensor-product space $\mathcal{H}$. 
    The Hamiltonian $\hat{H}$ is represented by a complex matrix $\mathbb{C}^{d^N \times d^N}$.
    In moderate dimensions, as considered throughout this article, the flow 
    \begin{equation}
        \hat \Phi^{\mathrm{SE}}_t = \exp\left( -t\frac{i}{\hbar}  \hat H \right)
    \end{equation}
    of Eq. \eqref{eq:SE} can be computed by evaluating the matrix exponential of $ -\frac{i}{\hbar} t \hat H$ numerically 
        \footnote{In the numerical experiments, the command {\tt exp} of the package {\tt LinearAlgebra.jl} is used.
        The underlying algorithm can exploit the Hermitian structure of matrices and is based on eigendecompositions.}.
    Approaches for high-dimensional systems may be found in Refs. \cite{Ceruti2024,Lubich2008}.
    Note that, for any $t$, the flow $\hat \Phi^{\mathrm{SE}}_t \colon \mathcal{H}\to \mathcal H$ is a unitary (thus, linear) operator.
    
    We can similarly compute solutions to the SSE \eqref{eq:SSE}.
    For each $k$ and any collection of states $|a_1\rangle,\ldots,|a_{k-1}\rangle,|a_{k+1}\rangle,\ldots,|a_N\rangle$, the flow $\hat\Phi^{\mathrm{SSE}}_{a_1,\ldots,a_{k-1},a_{k+1},\ldots,a_N,t} \colon \mathcal{H}_k \to \mathcal{H}_k$ of the $k$th separable Schrödinger equation
   
    can be computed as the matrix exponential of Hermitian matrices as
    \begin{equation}
        \hat\Phi^{\mathrm{SSE}}_{a_1,\ldots,a_{k-1},a_{k+1},\ldots,a_N,t}
        = \exp\left(-t\frac{i}{\hbar} \hat H_{a_1,\ldots,a_{k-1},a_{k+1},\ldots,a_N} \right).
    \end{equation}
    Next, we introduce an approximated flow map $\hat {\tilde \Phi}^{\mathrm{SSE}}_{\Delta t}$, with $\Delta t>0$ being a discretization parameter, of the  exact time-$\Delta t$-flow $\hat {\Phi}^{\mathrm{SSE}}_{\Delta t}$. 
    To compute this numerical $\hat {\tilde \Phi}^{\mathrm{SSE}}_{\Delta t}$, we consider the components of a given state $(|a_1^{[0]}\rangle,\ldots,|a_N^{[0]}\rangle) \in  \mathcal{H}^\times$, where upper indices in brackets $[\cdot ]$ are used to describe the effects of the update operations.
    Then, the updated components
    \begin{equation}
        (|a_1^{[N]}\rangle,\ldots,|a_N^{[N]}\rangle) 
        =\hat {\tilde \Phi}^{\mathrm{SSE}}_{\Delta t}(|a_1^{[0]}\rangle,\ldots,|a_N^{[0]}\rangle)
    \end{equation}
    are computed by performing $N$ update operations,
    \begin{equation}
    \label{eq:sp1}
        \begin{split}
            &|a_1^{[l]}\rangle = |a_1^{[l-1]}\rangle,\; \ldots,\;
            |a_{l-1}^{[l]}\rangle = |a_{l-1}^{[l-1]}\rangle\\
            &|a_{l}^{[l]}\rangle = \hat\Phi^{\mathrm{SSE}}_{a_1^{[l-1]},\ldots,a_{l-1}^{[l-1]},a_{l+1}^{[l-1]},\ldots,a_N^{[l-1]},\Delta t} |a_{l}^{[l-1]}\rangle\\
            &|a_{l+1}^{[l]}\rangle = |a_{l+1}^{[l-1]}\rangle,\; \ldots,\;
            |a_{N}^{[l]}\rangle = |a_{N}^{[l-1]}\rangle,
         \end{split}
     \end{equation}
    for $l\in\{1,\ldots,N\}$.
    In other words, the components are updated sequentially by the flow of the corresponding reduced operator;
    that is, in Eq. \eqref{eq:SSE}, the first component is updated while all other components remain unchanged, then the second component is updated while all other components remain unchanged, etc.

    To an initial value $(|a_1^{(0)}\rangle,\ldots,|a_N^{(0)}\rangle) \in \mathcal{H}^\times$, where we denote iterates with upper indices in parenthesis $(\cdot)$, the iteration
    \begin{equation}
    \label{eq:sp2}
        (|a_1^{(j+1)}\rangle,\ldots,|a_N^{(j+1)}\rangle)
        = \hat {\tilde \Phi}^{\mathrm{SSE}}_{\Delta t}(|a_1^{(j)}\rangle,\ldots,|a_N^{(j)}\rangle), 
    \end{equation}
    for $j\in\{0,1,\ldots\}$, constitutes a numerical approximation of first order $\mathcal{O}(\Delta t)$ of the exact solution of Eq. \eqref{eq:SSE}. 
    Both, the exact flow at time $\Delta t$, $\hat {\Phi}^{\mathrm{SSE}}_{\Delta t}$, and the numerical flow map $\hat {\tilde \Phi}^{\mathrm{SSE}}_{\Delta t}$ are nonlinear, yet inner-product-preserving (thus, norm-preserving) transformations, generalizing unitary maps from the SE to the flow of the nonlinear SSE. 
    In this sense, the numerical method is structure-preserving.

    The above numerical method is known as a splitting method \cite{Blanes_Casas_Murua_2024}, wherein each summand in the split is obtained by setting all but one right-hand side in Eq. \eqref{eq:SSE} to zero.
    Indeed, $\hat {\tilde \Phi}^{\mathrm{SSE}}_{\Delta t}$ amounts to computing the Lie-Trotter splitting, a first-order numerical method. 
    Higher-order splittings require more complicated compositions of the flow maps of each summand:
    for instance, we can perform a second-order numerical method $\hat {\doubletilde{$\Phi$}}^{\mathrm{SSE}}_{\Delta t}$ for approximating ${\hat \Phi}^{\mathrm{SSE}}_{\Delta t}$ by the Strang-Splitting method
    \begin{equation}\label{eq:StrangSSE}
        (|a_1'''\rangle,|a_2'''\rangle) = \hat {\doubletilde{$\Phi$}}^{\mathrm{SSE}}_{\Delta t}(|a_1\rangle,|a_2\rangle),    
    \end{equation}
    for the example $N=2$ and with
    \begin{equation}
        \begin{split}
            |a_1'\rangle &= |a_1\rangle\\
            |a_2'\rangle &= \hat\Phi^{\mathrm{SSE}}_{a_1,\Delta t/2}|a_2\rangle \\
            |a_1''\rangle &= \hat\Phi^{\mathrm{SSE}}_{a_2,\Delta t}|a_1\rangle\\
            |a_2''\rangle &= |a'_2\rangle \\
            |a_1'''\rangle &= |a''_1\rangle\\
            |a_2'''\rangle &= \hat\Phi^{\mathrm{SSE}}_{a_1,\Delta t/2}|a''_2\rangle.
            \end{split}
    \end{equation}
    By construction, $\hat {
    \doubletilde{$\Phi$}}^{\mathrm{SSE}}_{\Delta t}$ in Eq. \eqref{eq:StrangSSE} is a unitary-like, nonlinear transformation on the Cartesian product space $\mathcal{H}_1 \times \mathcal H_2$. 
    For higher-order splitting methods, as well as the non-bipartite case, see the literature on numerical splitting methods \cite{Blanes_Casas_Murua_2024,DAmbrosio2023,LeimkuhlerReich2005,McLachlan2002,HLW2013}.
    For splitting methods where the operators are unbounded, we additionally refer to Ref. \cite{ISERLES202429}.
    
    For any splitting method, when the numerical iterates $(|a_1^{(j)}\rangle,\ldots,|a_N^{(j)}\rangle)$ are mapped to $|\psi^{(j)}\rangle  = |a_1^{(j)} \rangle\otimes \ldots \otimes |a_N^{(j)}\rangle \in \mathcal{H}$, the norm $\sqrt{\langle \psi^{(j)}|\psi^{(j)}\rangle}$ remains constant.
    Moreover, for a Hamiltonian that decomposes into local parts,
    \begin{equation}
    \label{eq:DecomposedH}
            \hat H =
             \hat H_1 \otimes \hat 1_{\mathcal H_2} \otimes \ldots \otimes \hat 1_{\mathcal H_N}
            +\hat 1_{\mathcal H_1} \otimes \hat H_2 \otimes \hat 1_{\mathcal H_3} \otimes  \ldots \otimes \hat 1_{\mathcal H_N}  
            + \ldots
            +\hat 1_{\mathcal H_1} \otimes  \ldots \otimes \hat 1_{\mathcal H_{N-1}} \otimes \hat H_N,     
    \end{equation}
    the SSE \eqref{eq:SSE} decouple \cite{SW17}.
    Thus, a numerical solution based on a splitting method coincides with the exact solution because each individual equation in \eqref{eq:SSE} is solved exactly by the numerical scheme when the equations decouple.
    
    In the remainder of this section, we apply the method in Eqs. \eqref{eq:sp1} and \eqref{eq:sp2} to different Hamiltonians, comparing the solutions for the SSE and the SE.
    The source code for the numerical experiments may be found on GitHub \cite{GithubProject}.
    For convenience, we set $\hbar =1$.
    We show how, for small enough intervals, the numerical method renders the exact solution---in cases they are known---up to second order in $\Delta t$. 
    Therefore, this constitutes an interesting tool for witnessing dynamical entanglement without exactly solving the restricted dynamics.
    
    \subsection{Exchange interaction}
    \label{sec:NumericsSwapOperator}
    
        As a first example, we consider the two-qubit case $(N=2=d)$, with $a=a_1$, $b=a_2$, $\mathcal{H}_a = \mathcal{H}_1$, $\mathcal{H}_b = \mathcal{H}_2$, and the Hamiltonian performing a swap operation, $\hat H (| a \rangle \otimes | b \rangle) = | b \rangle \otimes | a \rangle$. 
        Consider an initial state $|a^0 \rangle  \otimes |b^0 \rangle \in \mathcal{H}_a \times \mathcal{H}_b$ with normalization $\langle a^0|a^0 \rangle = 1 = \langle b^0|b^0 \rangle$, the solution of the SE becomes
        \begin{align}\label{eq:SolSE}
		      |\psi(t)\rangle=\cos(t)|a^0\rangle \otimes |b^0\rangle-i\sin(t)|b^0\rangle \otimes |a^0\rangle.
        \end{align}
        For the SSE, the reduced operators read $\hat H_a = |a\rangle \langle a|$, $\hat H_b = |b\rangle \langle b|$, 
        and the exact solution of the SSE in Eq. \eqref{eq:SSE} can be derived from the reduced operators as \cite{SW17}
        \begin{equation}\label{eq:ExactSwap}
            \begin{split}
                |a(t)\rangle &= \cos(|q| t) |a^0\rangle - i \frac{q^\ast}{|q|} \sin(|q|t) |b^0\rangle\\
                |b(t)\rangle &=\cos(|q| t) |b^0\rangle - i \frac{q}{|q|} \sin(|q|t) |a^0\rangle,
            \end{split}
        \end{equation}
        which preserves this normalization and the transition amplitude $q = \langle a|b\rangle$. Here $q^\ast = \langle b|a\rangle$ denotes the complex conjugate of $q$.
        
        The Lie-Trotter splitting amounts to $\hat {\tilde \Phi}^{\mathrm{SSE}}_{\Delta t}(|a\rangle,|b\rangle) = (|a''\rangle,|b''\rangle)$ 
        \begin{equation}
            \begin{split}
                |a'\rangle &= \exp\left( -i \Delta t |b\rangle \langle b| \right)|a\rangle\\  
                |b'\rangle &= |b\rangle\\
                |a''\rangle &= |a'\rangle = \exp\left( -i \Delta t |b\rangle \langle b| \right)|a\rangle\\
                |b''\rangle &= \exp\left( -i \Delta t |a'\rangle \langle a'| \right)|b\rangle 
            \end{split}
        \end{equation}
        Using the series expansion of the matrix exponential map and $(|a\rangle\langle a|)^k = |a\rangle\langle a|$ for $k \in\{1,2,\ldots\}$, this can be simplified further to
        \begin{equation}\label{eq:LieTrotterSwap}
            \begin{split}
                |a''\rangle &= |a\rangle +  q^* \left( e^{-i \Delta t}-1\right) |b\rangle\\
                |b''\rangle &= q\left(1- e^{i \Delta t}\right) |a \rangle + \left(1+2|q|^2\left(-1+\cos(\Delta t)\right)\right)|b\rangle.
            \end{split}    
        \end{equation}
        This solution coincides with the analytical solution in Eq. \eqref{eq:ExactSwap} at time $\Delta t$ up to terms of order $\Delta t^2$ as the method is of first order.
        Like the exact solution, the method conserves $q=\langle a|b \rangle$ and the normalization of $|a\rangle$ and $|b\rangle$.
        
\begin{figure}
    \centering
    \includegraphics[ width=0.4\linewidth]{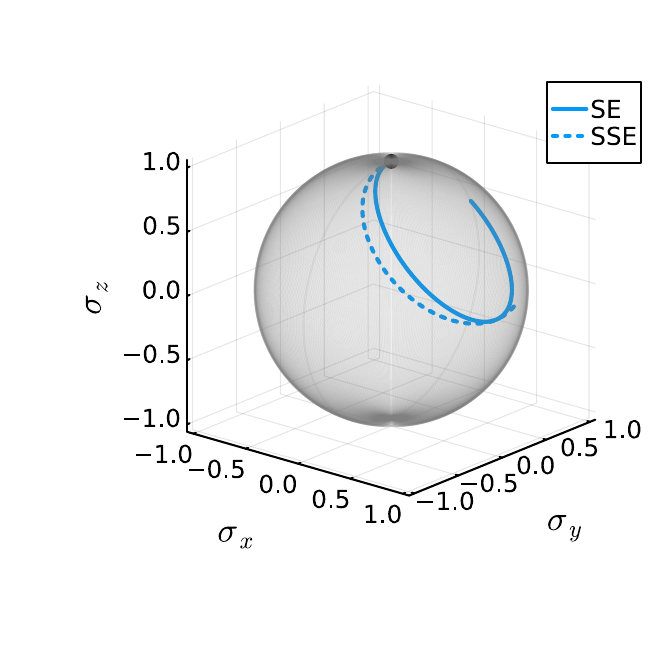}
    \includegraphics[ width=0.4\linewidth]{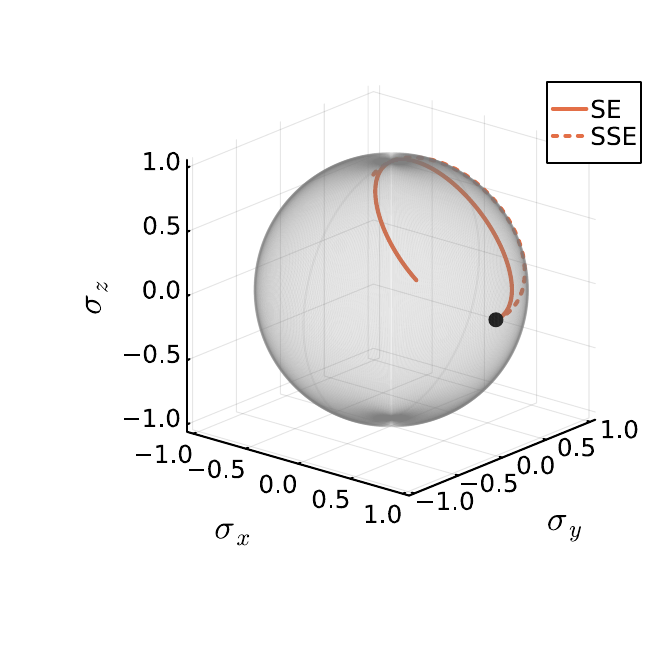}
    \caption{%
        Trajectories on the Bloch sphere of restricted (dashed) and non-restricted dynamics (solid) for $\Delta t=0.001$ and the initial state $|\psi(0)\rangle=|0\rangle\otimes(|0\rangle+|1\rangle)/\sqrt{2}$ for an exchange interaction (left, first qubit; right, second qubit).
    }\label{fig:ExchangeInteraction}
\end{figure}

        In \cref{fig:ExchangeInteraction}, we compare the trajectories for the restricted (SSE) and unrestricted (SE) evolutions of the system under the exchange interaction Hamiltonian. 
        The discretization parameter $\Delta t=0.001$ was chosen so that the exact and numerical solutions for the restricted evolution are visually indistinguishable in the plots.
        The figure shows the evolution of each subsystem after the partial trace over the other subsystem in the Bloch-sphere representation. 
        The restricted evolution renders a separable state for all times, and therefore, the subsystem state is pure, and the trajectory remains on the surface of the sphere. 
        By contrast, for the unrestricted evolution, the partial trace produces a mixed state, and the trajectory travels through the interior of the sphere.
        
        We highlight that the numerical approach converges to the analytical solutions for small enough time steps.
        Moreover, backward error analysis \cite{HLW2013} can be applied to analyze the Trotter scheme further, as well as to analzse the Strang-Splitting;
        the corresponding results are provided in Appendix \ref{sec:AppendixBEA}.
        
    \subsection{Random Hamiltonian}
        
\begin{figure}
    \centering
    \includegraphics[trim=1.4cm 0cm 0cm 0cm, clip,width=0.33\linewidth]{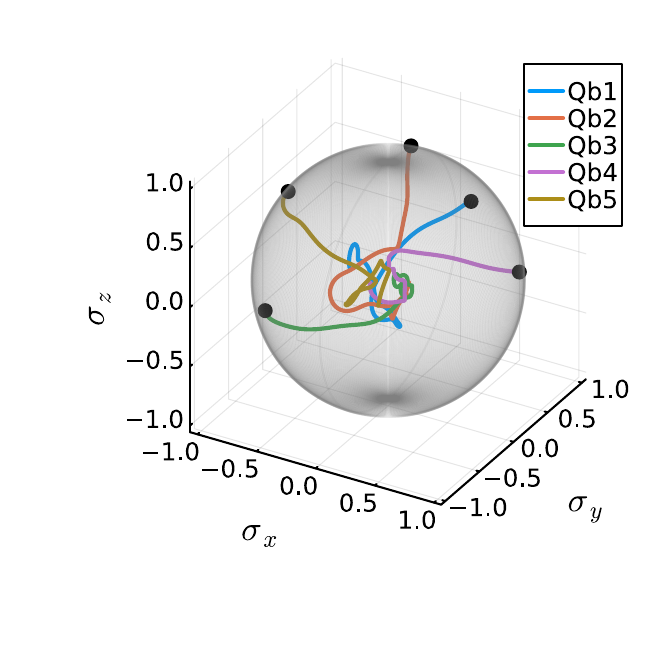}
    \includegraphics[trim=1.4cm 0cm 0cm 0cm, clip,width=0.33\linewidth]{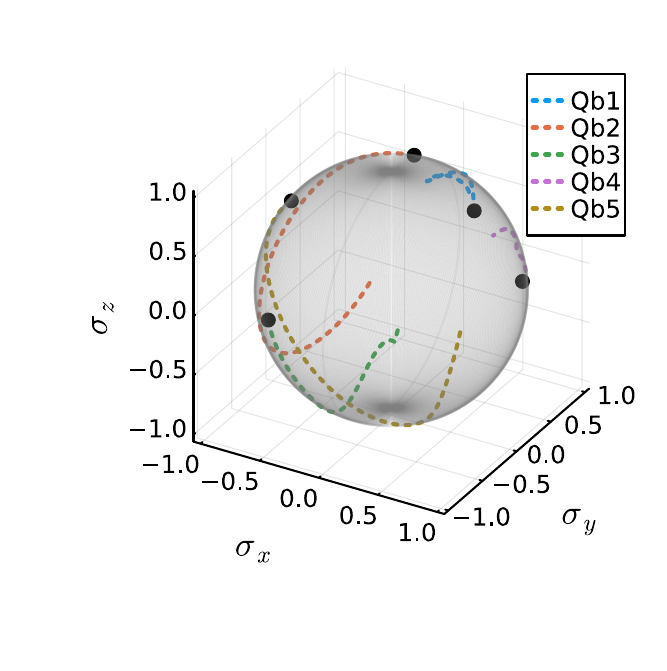}
    \includegraphics[trim=1.4cm 0cm 0cm 0cm, clip,width=0.33\linewidth]{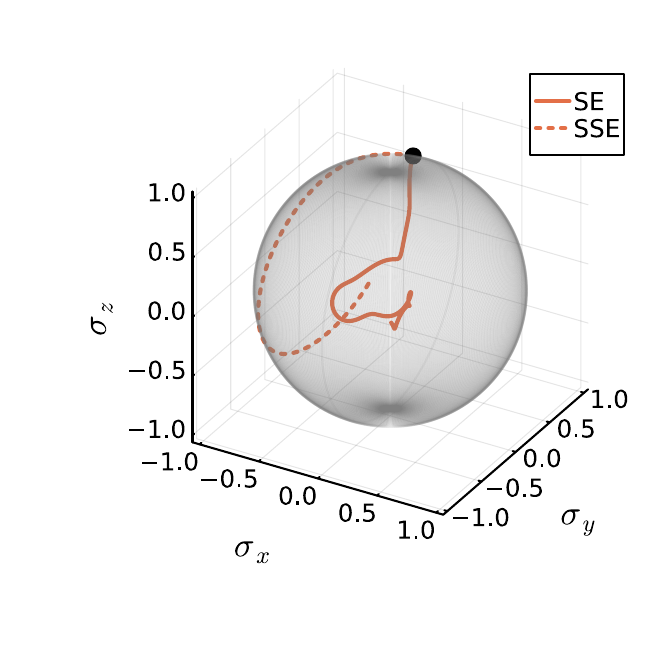}
    \includegraphics[width=0.35\linewidth]{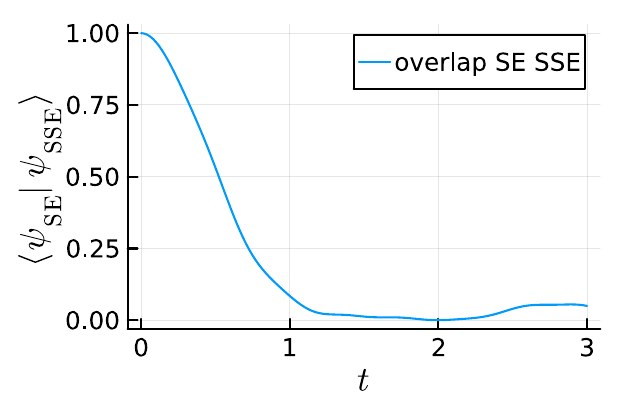}
    \includegraphics[width=0.35\linewidth]{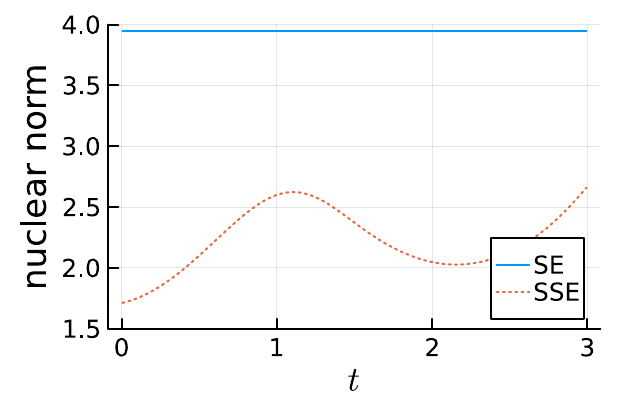}
    
    \caption{%
        Top: Bloch sphere with the evolution of the qubits (Qb1--Qb5) generated by a random Hamiltonian (left: SE \eqref{eq:SE}, middle: SSE \eqref{eq:SSE}, right: comparison between trajectories for the second qubit). In the bottom,  the left panel depicts the overlap $\langle \psi_{\mathrm{SE}}| \psi_{\mathrm{SSE}}\rangle$.
        The right panel shows the
        rate of change of the SSE, $\big| \frac{d}{d t} | \psi_{\mathrm{SSE}}(t)\rangle \langle \psi_{\mathrm{SSE}}(t)| \big|_{\mathrm{nucl}}$ (dotted line), and the corresponding quantity for the SE (solid line), using the nuclear/trace norm.
    }\label{fig:RandHOperatorDyn}
\end{figure}

        We now consider $N=5$ qubits 
        and use a randomly sampled Hermitian operator as a Hamiltonian.
        For this, 528 real and imaginary parts of complex numbers are sampled independently from a normal distribution to form the free entries of a $2^5 \times 2^5$ Hermitian matrix $\hat H$.
        In 
        \cref{fig:RandHOperatorDyn}, the trajectory of each subsystem is shown on a Bloch sphere, 
        demonstrating that the SE and SSE dynamics $\Psi^{\mathrm{SE}}$, $\Psi^{\mathrm{SSE}}$ can look very different, highlighting the role of entanglement in such dynamics.
        In particular, we can see how the partial trace for obtaining the Bloch-sphere representations brings trajectories evolving on average to the maximally mixed state (top-left plot in \cref{fig:RandHOperatorDyn}). 
        This shows that the unrestricted trajectory brings the full system to a maximally entangled state, in stark contrast with the restricted evolution that uses the same Hamiltonian but cannot produce entanglement, Fig. \ref{fig:RandHOperatorDyn}  top-middle plot.
        Specifically, \cref{fig:RandHOperatorDyn} (bottom-left) shows that the overlaps between the separable and inseparable solutions rapidly decay.
        Also, the speed of the entangling (SE) and non-entangling (SSE) evolution yields different rates of changes of the state, as seen in the  \cref{fig:RandHOperatorDyn} (bottom-right) via the nuclear (trace) norm.
        
    \subsection{Ladder operator}

        As a final example in this section, we study the case of two-party correlations between three qudits $(N=3=d)$.
        Let us start by considering a complex Hilbert space $\mathcal{H}_J$ with three basis elements labeled as $|-1\rangle, |0\rangle, |1\rangle$, relating to systems of angular momentum.
        The ladder operators for the angular-momentum quantum number $j=1$
        are defined as the linear extension of 
        \begin{equation}
            \begin{split}
                \hat J_+ |-1 \rangle &= \sqrt{2}|0 \rangle, \quad
                \hat J_+ |0 \rangle = \sqrt{2}|1 \rangle, \quad
                \hat J_+ |1 \rangle = 0\\
                \hat J_- |-1 \rangle &= 0, \quad
                \hat J_- |0 \rangle = \sqrt{2}|-1 \rangle, \quad
                \hat J_- |1 \rangle = \sqrt{2}|0 \rangle.
            \end{split}
        \end{equation}
        The full Hilbert space considered is $\mathcal{H} = (\mathcal{H}_J)^{\otimes 3}$, and the Hamiltonian
        \begin{equation}
            \hat H = \sum_{k_1,k_2,k_3}\big( \eta_{k_1k_2k_3}\hat J_+^{k_1} \otimes J_+^{k_2}\otimes J_+^{k_3}
            +\eta_{k_1k_2k_3}^*\hat J_-^{k_1} \otimes J_-^{k_2}\otimes J_-^{k_3}\big),
        \end{equation}
        where ${k_1,k_2,k_3 \in \{0,1\}}$ and $\eta =(\eta_{k_1k_2k_3})_{0\le k_1,k_2,k_3\le 1} \in \mathbb{C}^{2\times 2 \times 2}$ are coupling strength parameters.
        We consider $\eta$ with
        \begin{equation}
            \eta_{k_1k_2k_3} = 
            \begin{cases}
                1, \quad k_1+k_2+k_3=r\\
                0, \quad \text{otherwise},
            \end{cases}
        \end{equation}
        which corresponds to an $r$-party correlator. 
        With this method, we can discriminate and study both the two-party and three-party correlation cases,
        \begin{equation}
            \hat H =  \eta_{100}\hat J_+\otimes\hat 1_{\mathcal H_2} \otimes \hat 1_{\mathcal H_3}+\eta_{010} \hat 1_{\mathcal H_1} \otimes\hat J_+ \hat\otimes \hat 1_{\mathcal H_3}+\eta_{001} \hat 1_{\mathcal H_1} \otimes \hat 1_{\mathcal H_1}\otimes\hat J_+
            +\text{h.c.}
            \quad\mathrm{and}\quad
            \hat H \propto \hat J_+ \otimes J_+\otimes J_++\text{h.c.},
        \end{equation}
        respectively.

\begin{figure}
    \centering
    \includegraphics[trim=0cm 0cm 0cm 0cm, clip,width=0.33\linewidth]{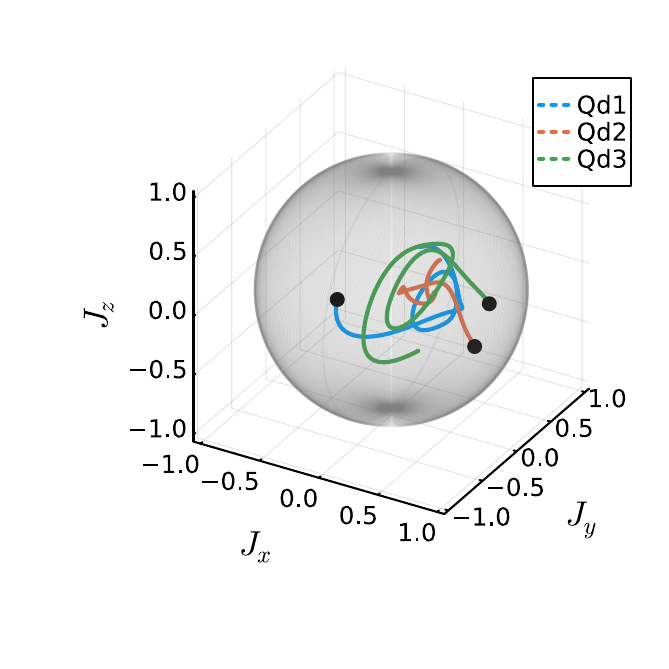}\hfill
    \includegraphics[trim=0cm 0cm 0cm 0cm, clip,width=0.33\linewidth]{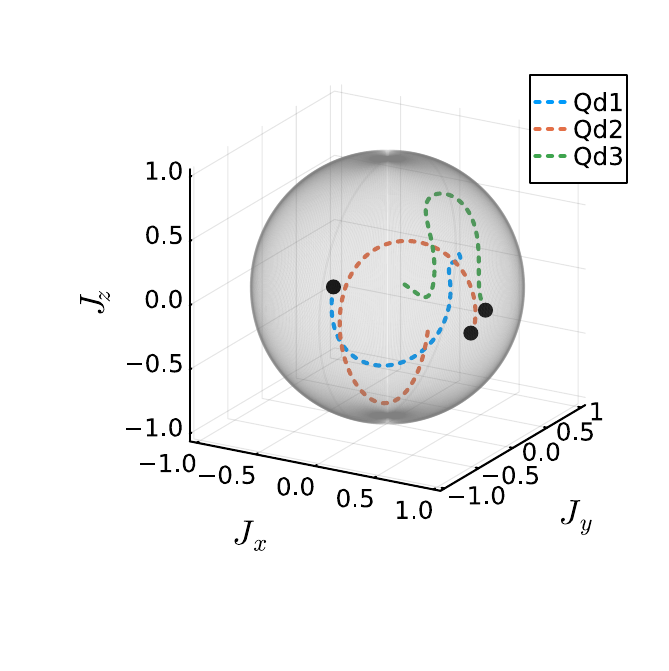}\hfill
    \includegraphics[trim=0cm 0cm 0cm 0cm, clip,width=0.33\linewidth]{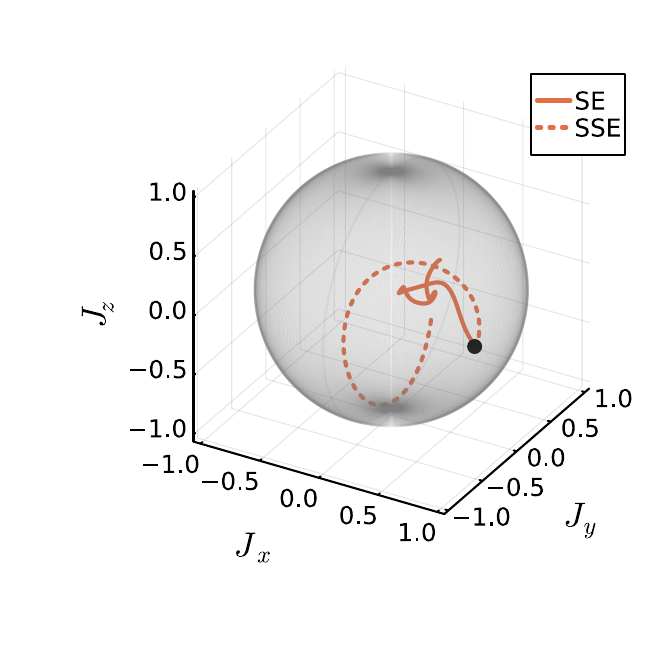}
    \includegraphics[width=0.35\linewidth]{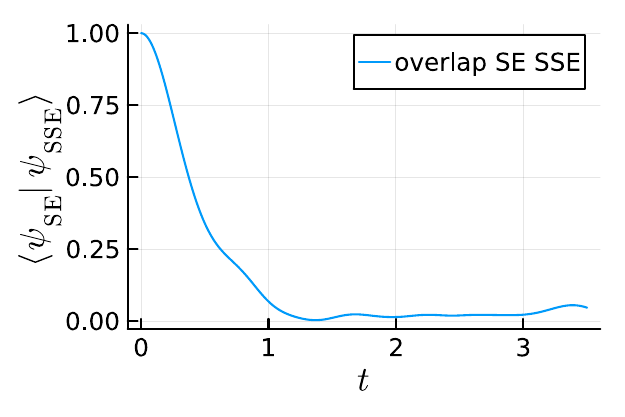}
    \includegraphics[width=0.35\linewidth]{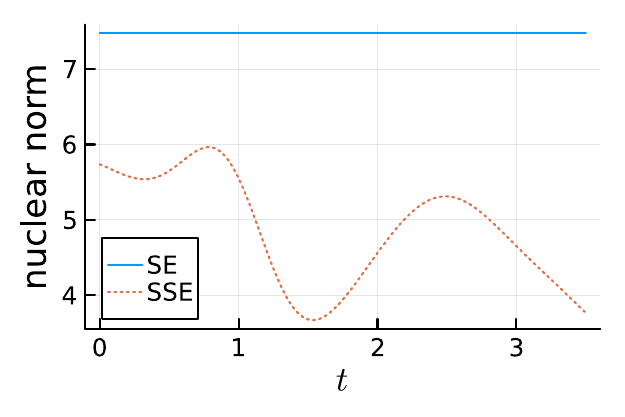}
    \caption{%
        Top: Dynamics based on a Hamiltonian given in terms of two-party correlators of angular-momentum ladder operators.
        The evolution of qudits is shown on the Poincar{\'e} sphere (left: SE, middle: SSE, right: comparison for second qudit).
        Bottom: the overlap $\langle \psi_{\mathrm{SE}}| \psi_{\mathrm{SSE}}\rangle$ (left), as well as the rate of change $\big| \frac{d}{d t} | \psi_{\mathrm{SSE}}(t)\rangle \langle \psi_{\mathrm{SSE}}(t)| \big|_{\mathrm{nucl}}$ in nuclear norm for the SSE (dotted line, right plot) and for the SE (solid line, right plot) as a function of time.
    }\label{fig:LadderOperator1}
\end{figure}

\begin{figure}
    \centering
    \includegraphics[trim=0 0cm 0cm 0cm, clip,width=0.31\linewidth]{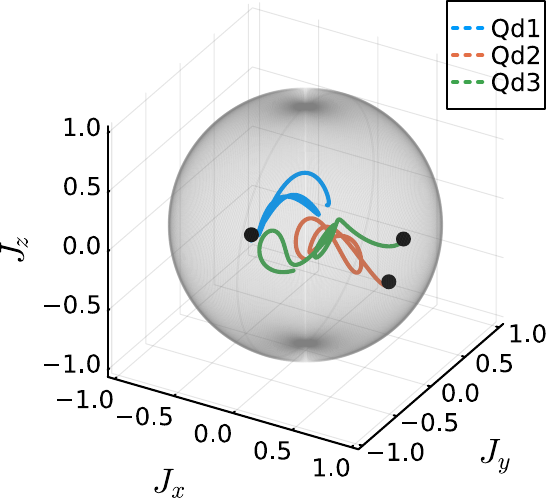}\hfill
    \includegraphics[trim=0 0cm 0cm 0cm, clip,width=0.31\linewidth]{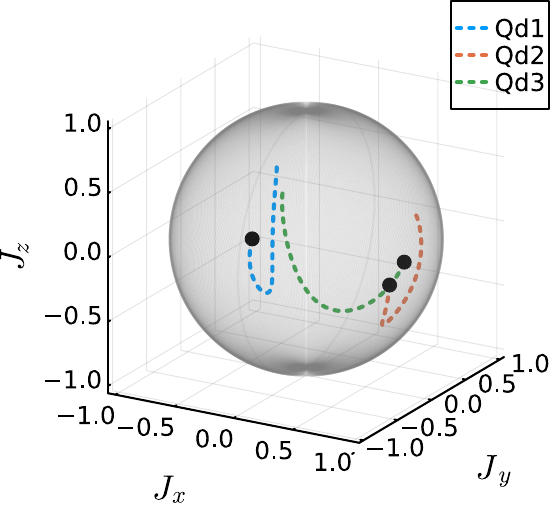}\hfill
    \includegraphics[trim=0 0cm 0cm 0cm, clip,width=0.31\linewidth]{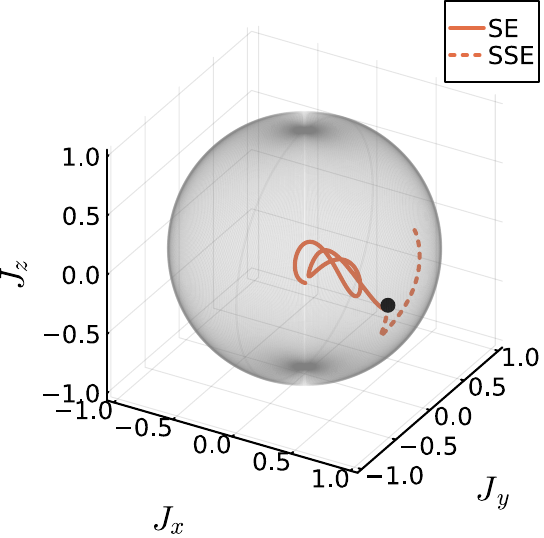}
    \includegraphics[width=0.40\linewidth]{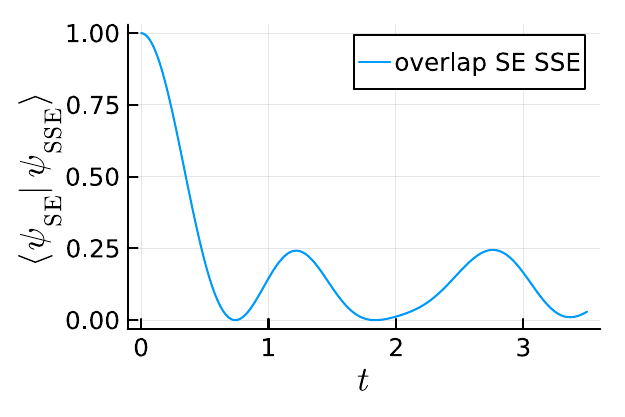}
    \includegraphics[width=0.40\linewidth]{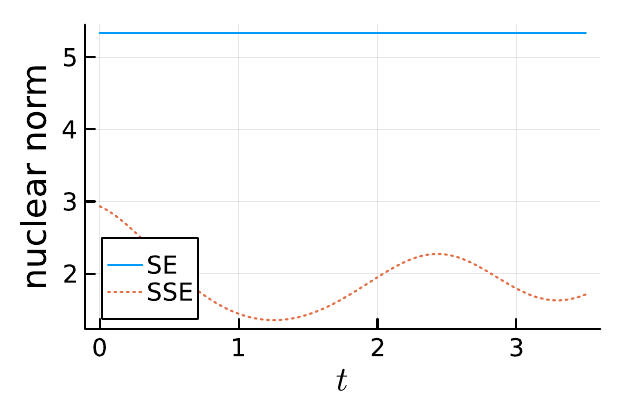}
    \caption{%
        Top: Dynamics via angular-momentum ladder operator and $3$-party correlator on the  Poincar{\'e} sphere (left: SE, middle: SSE, right: comparison second qudit).
        Bottom: Overlap of the restricted (SSE) and unrestricted (SE) dynamics (left panel), and the individual rates of change (i.e., speed; right panel).
        Compared to the two-party correlator in \cref{fig:LadderOperator1}, the here-discussed three-party correlator yields relatively strong revivals (increases of the overlap) in the similarity of the SE and SEE solutions.
    }\label{fig:LadderOperator3}
\end{figure}

        Dynamics for a 2-party correlator are displayed in \cref{fig:LadderOperator1}. 
        \Cref{fig:LadderOperator3} display the same experiments for a $3$-party correlator.
        The motion is visualized on the Poincar{\'e} sphere as the higher-dimensional analog to the Bloch sphere.
        In both cases, we see how the trajectories for the reduced systems disagree as a clear signature of quantum correlations of entanglement caused by the interaction. 
        The difference between the two types of correlators can be seen in the overlap with the unrestricted trajectory, where we can see certain revivals for the 3-party correlator, absent in the 2-party correlator, at least on this time scale.
    
    In the first set of results we discussed in this section, we have used splitting methods to numerically obtain and compare linear and unrestricted SE solutions with nonlinear and restricted SEE solutions.
    Thereby, we showed that interesting types of entangling dynamics can be characterized.
    Using known exact solutions, we first benchmarked the numerical approach;
    using a randomly generated five-qubit Hamiltonian, we secondly demonstrated the application to arbitrary multi-qubit interactions.
    Thirdly, multi-party correlators beyond qubits were used to exemplify a study of distinct types of interactions and their impact on the entangling power of processes that couple angular momenta of three particles in different ways.

    In the next section, we change our methodology significantly.
    Rather than using the equations of motion derived from the principle of stationary actions and then formulating numerical solvers, we can introduce a more direct approach. 

\section{Least-action-based numerical approach to restricted quantum systems}

The methodology of numerical integration of SSE presented in \cref{sec:NumericsSplitting} does not inherently exploit that the SSE were derived from a variational principle.
While numerical splitting methods constitute a natural approach to integrate the SSE \eqref{eq:SSE}, their successful application requires that continuous SSE are derived analytically first, before numerical experiments can be carried out.
This becomes particularly challenging when notions of quantumness other than entanglement are considered \cite{SW20}.

To avoid this initial step, we propose an approach that allows us to compute the motion of restricted quantum systems without the need to derive equations of motion analytically first. 
For this purpose, we discretize on the level of variational principles already to obtain variational integration methods \cite{MarsdenWest2001,EllisonEtAl2018,OberBloebaum2015} for the restricted dynamics.
These methods are structure-preserving by design;
i.e., the numerical solution obeys a discrete variational principle.
This approach follows the philosophy of geometric numerical integration \cite{HLW2013,LeimkuhlerReich2005,DAmbrosio2023}: a numerical solution should share the geometric properties of the continuous system in order to facilitate robust and reliable numerical computations, avoiding numerical artifacts.
In addition, this method allows practitioners to easily experiment with various restrictions.

In this section, we follow this path and implement a variational discretization for Lagrangians that are linear in velocities. 
Then, we compare the results with the approach followed in the previous section for the exchange interaction Hamiltonian.
Also, we change notation to a more convenient one in the context studied here, such as $|\psi\rangle\to\Psi$ and $\langle\psi|\to \bar\Psi$.

\subsection{Variational discretizations for Lagrangians linear in velocities}
\label{sec:VarIntegrationSeparable}

    Consider the action functional
    \begin{equation}
    \label{eq:actionS}
        S(\Psi) = \int_{t_0}^{t_M}L(\Psi(t),\overline{\Psi}(t),\dot \Psi(t),\dot {\overline{\Psi}}(t)) dt    
    \end{equation}
    on the space of $\mathbb{C}^d$-valued, smooth curves $\Psi$ defined on an interval $[t_0,t_M]$.
    Above, $\overline{\Psi}(t)$ denotes the complex conjugate of ${\Psi}(t)$.
    Moreover, $^\top$ denotes the transposition of a vector and matrix
    \footnote{\label{foot:newNotation}
        In this section (\cref{sec:VarIntegrationSeparable}) and Appendix \ref{sec:AppendixProofCommutativeDiagram}, we refrain from using the bra-ket notation.
            This avoids the requirement to restrict ourselves to operations that descend to complex projective spaces, i.e., that are equivariant with respect to complex rescaling.
            Indeed, our discussion includes variational discretizations that may or may not have this property.
        }.
    Consider, in particular, the Lagrangian function
    \begin{equation}
    \label{eq:L_SE}
        L(\Psi,\bar{\Psi},\dot \Psi,\dot {\bar{\Psi}}) = \frac{i \hbar}{2}\left(\bar{\Psi}^\top \dot \Psi  -  \dot {\bar{\Psi}}^\top  \Psi \right) -  \bar{\Psi}^\top  \hat H \Psi,
    \end{equation}
    defined for any $\Psi,\bar{\Psi},\dot \Psi,\dot {\bar{\Psi}} \in \mathbb C^d$. 
    We decorate a symbol by a bar, such as $\bar{\Psi}$, to denote input variables of functions into which we later substitute the complex conjugate $\overline \Psi$ of $\Psi$.  
    A curve $\Psi$ is stationary for the action $S$ in Eq. \eqref{eq:actionS} with respect to smooth variations that fix endpoints if and only \eqref{eq:SE} is fulfilled \cite{SW17,Lubich2008}.
    
    For a discretization parameter $\Delta t = t_M/M$ $(M \in \mathbb{N} \setminus \{0\})$ consider $t_j = t_0+j\Delta t$ $(j\in\{1,\ldots,M\})$.
    A discrete action $S_\Delta$ is defined on discrete curves modeled as vectors $\Psi_\Delta$ with $\Psi_\Delta = (\Psi^{(0)},\ldots,\Psi^{(M)}) \in \mathbb{C}^{d(M+1)}$.
    A discrete action $S_\Delta$ takes the form
    \begin{equation}
    \label{eq:Sdelta}
        S_\Delta(\Psi_\Delta) = \sum_{j=0}^M  L_\Delta (\Psi^{(j)},\overline \Psi^{(j)}, \Psi^{(j+1)},\overline \Psi^{(j+1)})     
    \end{equation}
    for a discrete Lagrangian $L_\Delta \colon (\mathbb{C}^{d})^4  \to \mathbb{C}$ that is analytic on a domain of interest. 
    Moreover, we require the following symmetry and reality property $L_\Delta (\Psi^{(j)},\bar \Psi^{(j)}, \Psi^{(j+1)},\bar \Psi^{(j+1)})= L_\Delta (\bar \Psi^{(j)}, \Psi^{(j)}, \bar \Psi^{(j+1)},\Psi^{(j+1)}) \in \mathbb{R}$, whenever $\bar \Psi^{(j)} = \overline \Psi^{(j)}$ and $\bar \Psi^{(j+1)} = \overline \Psi^{(j+1)}$, respectively.
    Further, we consider the exact discrete Lagrangian $L_\Delta^{\mathrm{ex}}$ given as
    \begin{equation}
        \begin{split}
            &L_\Delta^{\mathrm{ex}}(\Psi^{(j)},\bar \Psi^{(j)}, \Psi^{(j+1)},\bar \Psi^{(j+1)})  \\
            &=  \int_{t_j}^{t_{j+1}} L(\Psi(t),\bar{\Psi}(t),\dot \Psi(t),\dot {\bar{\Psi}}(t)) dt,        
        \end{split}
    \end{equation}
    where $\Psi(t)$ is the solution to the Euler--Lagrange equation to $L$ in Eq. \eqref{eq:SE}, with $\Psi(t_j)=\Psi^{(j)}$, $\bar \Psi(t_j)=\bar \Psi^{(j)}$ 
        \footnote{
            For general Lagrangians, when the Euler--Lagrange equations to $L$ are 2nd order differential equations (instead of first order as for the Lagrangian \eqref{eq:L_SE}) a 2-point boundary value problem $\Psi(t_j)=\Psi^{(j)}$, $\Psi(t_{j+1})=\Psi^{(j+1)}$,$\bar \Psi(t_j)=\bar \Psi^{(j)}$, $\bar \Psi(t_{j+1})=\bar \Psi^{(j+1)}$ is solved \cite{MarsdenWest2001}.
            }.
    On solutions of Eq. \eqref{eq:SE} and with $\Psi^{(j)} = \Psi(t_j)$, $\bar \Psi^{(j)} = \overline \Psi(t_j)$, the values $S(\Psi)=S_\Delta(\Psi_\Delta)$ coincide when $L_\Delta = L_\Delta^{\mathrm{ex}}$.
    For computations, the exact discrete Lagrangian $L_\Delta$ is approximated, for example, by
    \onecolumngrid
    \begin{equation}
            L^{\alpha}_\Delta(\Psi^{(j)},\bar{\Psi}^{(j)},\Psi^{(j+1)},\bar{\Psi}^{(j+1)}) = \Delta t L\left(\alpha \Psi^{(j)} + (1-\alpha) \Psi^{(j+1)},
            \alpha \bar{\Psi}^{(j)} + (1-\alpha) \bar{\Psi}^{(j+1)},
            \frac{\Psi^{(j+1)}-\Psi^{(j)}}{\Delta t},
            \frac{\bar{\Psi}^{(j+1)}-\bar{\Psi}^{(j)}}{\Delta t}
            \right),
        \end{equation}
        for $\alpha \in [0,1]$. 

        Variations of the discrete action w.r.t.~the interior points $(\Psi^{(1)},\ldots,\Psi^{(M-1)})$, $(\bar{\Psi}^{(1)},\ldots,\bar{\Psi}^{(M-1)})$
        yield the discrete Euler--Lagrange equations
        \begin{equation}
        \label{eq:DEL}
        \begin{split}
            &\phantom{+}\nabla_1 L_\Delta (\Psi^{(j)},\bar \Psi^{(j)}, \Psi^{(j+1)},\bar \Psi^{(j+1)})+
            \nabla_3 L_\Delta (\Psi^{(j-1)},\bar \Psi^{(j-1)}, \Psi^{(j)},\bar \Psi^{(j)}) =0\\
            &\phantom{+}\nabla_2 L_\Delta (\Psi^{(j)},\bar \Psi^{(j)}, \Psi^{(j+1)},\bar \Psi^{(j+1)})+\nabla_4 L_\Delta (\Psi^{(j-1)},\bar \Psi^{(j-1)}, \Psi^{(j)},\bar \Psi^{(j)}) =0
        \end{split}
        \end{equation}
        for $j\in\{1,\ldots,M-1\}$.
        Here, $\nabla_l L_\Delta $ $(l=1,2,3,4)$ denotes the complex derivative with respect to the $l$th input argument to $L_\Delta$.  
        As by the symmetry and reality property of $L_\Delta$, one of the equations in \eqref{eq:DEL} is sufficient already to describe the dynamics whenever $\bar{\Psi}^{(j)}$ is required to coincide with the complex conjugate $\overline{\Psi}^{(j)}$ of ${\Psi}^{(j)}$ for all $j\in\{0,\ldots,M\}$.
        This yields a two-term recursion in which 
        $\Psi^{(j+1)}$ can be computed from $\Psi^{(j-1)}$, $\Psi^{(j)}$ for $j\in\{1,\ldots,M-1\}$.
        To initiate the recursion given $\Psi^{(0)} \in \mathbb{C}^d$,
        notice that for any Lagrangian $L$ that is linear in $\dot \Psi$ and $\dot {\bar\Psi}$, the initial conjugate momentum $P^{(0)}=\frac{\partial L}{\partial \dot \Psi}(\Psi^{(0)},\overline \Psi^{(0)})$ can be computed from $\Psi^{(0)}$ only. Equating $P^{(0)}$ with the discrete initial momentum at $j=0$ yields 
        \begin{equation}
        \label{eq:InitStep}
            \frac{\partial L}{\partial \dot \Psi}(\Psi^{(0)},\overline{\Psi}^{(0)})
            = -\nabla_1 L_\Delta (\Psi^{(0)},\overline \Psi^{(0)}, \Psi^{(1)},\overline \Psi^{(1)}).
        \end{equation}
        \Cref{eq:InitStep} can be solved for $\Psi^{(1)}$. 
        The kinetic term $\bar{\Psi}^\top \dot \Psi  -  \dot {\bar{\Psi}}^\top  \Psi$ of the Lagrangian $L$ of \eqref{eq:L_SE} has a beneficial structure (linearity in $\Psi$ and $\dot \Psi$);
        that is, the scheme is numerically stable 
        \footnote{
            Stable in the sense of A-stability.
            This is a slight extension of a result in \cite{RowleyMarsden2002}. 
            For the notion of numerical stability, see, for instance, Ref. \cite{DAmbrosio2023}.
        }
        and of second order for $L_\Delta=L_\Delta^{\frac 12}$.
        It is numerically unstable for $L_\Delta=L_\Delta^{ 0}$ and $L_\Delta=L_\Delta^{ 1}$, for which the order of consistency is 1 \cite{kraus2017projectedvariationalintegratorsdegenerate,EllisonEtAl2018,RowleyMarsden2002}.

    \subsection{Two variational approaches to integrate restricted variational principles}
    \label{sec:VariationalIntegratorSSE}
        
\begin{figure*}
    \centering

    \begin{tikzcd}[sep=small]
    & {S(\Psi)=\int L(\Psi,\dot \Psi) dt)} \arrow[ld, "\Psi=a \otimes b" description] \arrow[rd, "\Delta" description] &                                                                         \\
    {
        \begin{array}{c}
        S^{\mathrm{sep}}(a,b) \\= \int \underbrace{L(a\otimes b, \dot a \otimes b + a \otimes \dot b)}_{= L^{\mathrm{sep}}((a,b),(\dot a,\dot b))} d t
        \end{array}		
        } \arrow[rd, "\Delta" description] &                                                                                                                  &
        \begin{array}{c}
            S_d((\Psi^{(i)})_i)\\
            =\sum_i \Delta t L_\Delta(\Psi^{(i)},\Psi^{(i+1)})
        \end{array}		
        \arrow[ld, "\Psi = a \otimes b" description] \\
    & {
        \begin{array}{c}
        S_\Delta^{\mathrm{sep}1}(a^{(i)},b^{(i)})_i
        = \sum_i \Delta t L_\Delta^{\mathrm{sep}} ((a^{(i)},b^{(i)}),(a^{(i+1)},b^{(i+1)}))\\
        \neq\\
        S_\Delta^{\mathrm{sep}2}(a^{(i)},b^{(i)})_i =
        \sum_i \Delta t L_\Delta(a^{(i)}\otimes b^{(i)},a^{(i+1)}\otimes b^{(i+1)})
    \end{array}
        }     &                                                                        
    \end{tikzcd}   
    \caption{%
        Visualization of the two distinct approaches: ``first-restrict-then-discretize" (left path) and ``first-discretize-then-restrict" (right path).
        For compactness of notation, we show the bipartite case $N=2$ with variables $a,b$ and neglect the complex conjugate input arguments of the Lagrangians.
        ``$\xrightarrow{\Delta}$'' denotes discretization by a variational integrator. The diagram does not commute as the left branch yields $S^{\mathrm{sep1}}$ and the right branch $S^{\mathrm{sep2}}$.
    }\label{fig:DiagramVarIntegrator}
\end{figure*}
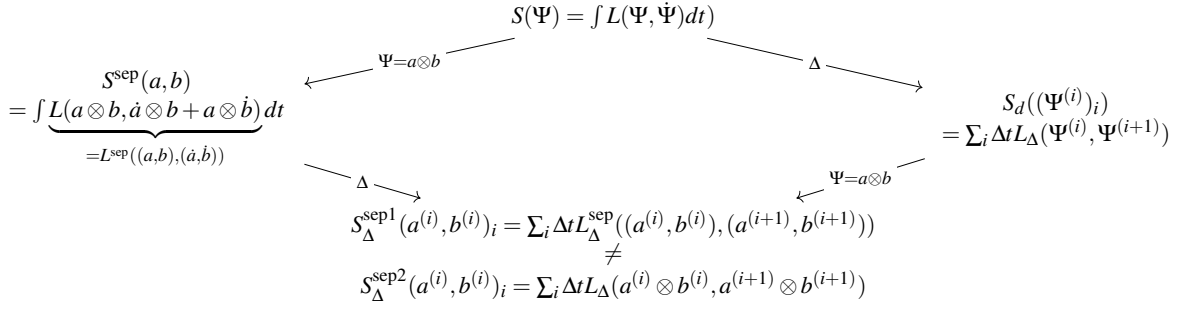
        
        There are two natural ways to apply variational, numerical integration schemes to restricted variational principles. 
        The two approaches are compared in \cref{fig:DiagramVarIntegrator}.
        Starting from the continuous, full variational principle for the action $S$, 
        \begin{enumerate}
            \item 
            we can restrict the continuous action $S(\Psi)$ to separable curves, obtaining the action
            \begin{equation}
                S^{\mathrm{sep}}(a_1,\ldots,a_N) = S(a_1 \otimes \ldots \otimes a_N), 
            \end{equation}
            which is defined on $\mathcal{H}^\times$-valued curves. 
            Then, a variational integrator can be applied to $S^{\mathrm{sep}}$. 
            We can summarize the approaches as a ``first-restrict-then-discretize'' technique (left path in \cref{fig:DiagramVarIntegrator}).
            
            \item
            We can discretize the continuous, full variational principle $S$ to form a discrete action $S_\Delta$ of the form \eqref{eq:Sdelta}. 
            Then, we restrict the discrete action $S_\Delta\colon \mathcal{H} \to \mathbb{R}$ to separable curves;
            i.e., we form $S_\Delta^{\mathrm{sep}}\colon (\mathcal{H}^\times)^{M+1} \to \mathbb{R}$ with
            \begin{equation}
            S_\Delta^{\mathrm{sep}}\left((a^{(j)}_1,\ldots,a_N^{(j)})_{j=0}^M\right)
                = S_\Delta\left((a_1^{(j)} \otimes \ldots \otimes a_N^{(j)})_{j=0}^M\right).
            \end{equation}
            Then, the discrete Euler-Lagrange equations are computed.
            We can summarize the approach as a ``first-discretize-then-restrict'' method (right path in \cref{fig:DiagramVarIntegrator}).
        \end{enumerate}
        Note that the first approach is particularly easy to implement \cite{GithubProject}:
        when a variational integration scheme is implemented that takes a general Lagrangian $L$ as its input, this algorithm can simply be supplied with the restricted Lagrangian, $L^\mathrm{sep}$, i.e., with 
        \begin{equation}
                L^\mathrm{sep}\left(
                    \begin{pmatrix}
                    a_1\\ \vdots \\a_N
                    \end{pmatrix},
                    \begin{pmatrix}
                    \bar{a_1}\\ \vdots \\\bar{a_N}
                    \end{pmatrix},
                    \begin{pmatrix}
                    \dot a_1\\ \vdots \\ \dot a_N
                    \end{pmatrix},
                    \begin{pmatrix}
                    \dot{\bar{a_1}}\\ \vdots \\ \dot{\bar{a_N}}
                    \end{pmatrix}
                \right)
                =L\left(
                    \bigotimes_{i=1}^N a_i,
                    \bigotimes_{i=1}^N \bar {a_i},
                    \sum_{j=1}^N 
                    \bigotimes_{i=1}^{j-1} a_i \otimes \dot a_j \otimes \bigotimes_{i=j+1}^N a_i,
                    \bigotimes_{i=1}^{j-1} \bar {a_i} \otimes \dot {\bar{a_j}}\otimes \bigotimes_{i=j+1}^N \bar {a_i}
                \right).
        \end{equation}
        This approach generalizes from separability conditions to similar variational restrictions.
        One can, therefore, compute motions to restricted variational problems numerically without analytically working out equations of motion to the restricted variational principle.
        
\begin{figure}
    \centering
    \includegraphics[width=0.45\linewidth]{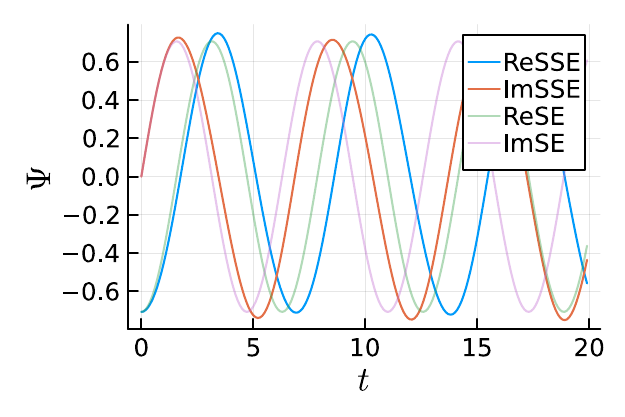}
    \includegraphics[width=0.45\linewidth]{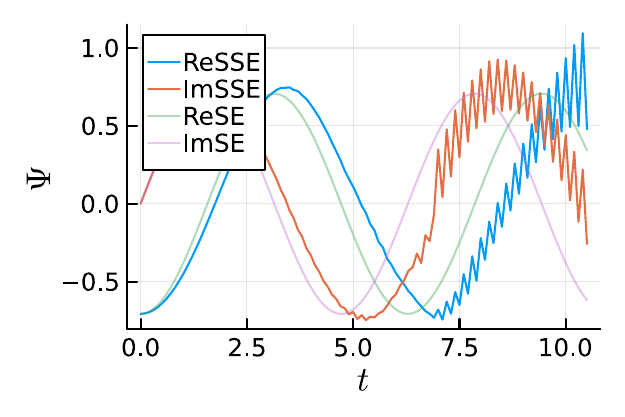}
    \caption{%
        Numerical integration of the separable dynamics computed with the two approaches of \cref{sec:VariationalIntegratorSSE}.
        Real and imaginary part of one component of $\Psi$ for the restricted (SSE) and unrestricted (SE) are shown as an example.
        The left depicts the ``first-restrict-then-discretize'' approach, and the right one is based on ''first-discretize-then-restrict''.
    } \label{fig:VarIntegrationSSE}
\end{figure}
        
        \Cref{fig:VarIntegrationSSE} shows an application of the two approaches, ``first-restrict-then-discretize'' and ``first-discretize-then-restrict'' for integrating the SSE for the swap operator (see \cref{sec:NumericsSwapOperator}) with time-discretization $\Delta t=0.1$ and the variational midpoint rule.
        The ``first-restrict-then-discretize" approach yields a stable numerical scheme as in the restricted Lagrangian the kinetic term (i.e., any summand containing $\dot a_j$, $\dot {\bar{a_j}}$, $j\in\{1,2\}$) is of the form $\eta(q) \cdot \dot q$, where $q=\begin{pmatrix}
        a_1,&  a_2,&  {\bar{a_1}},&{ \bar{a_2}}
        \end{pmatrix}^\top$, with $\eta$ linear and matrix valued \cite{kraus2017projectedvariationalintegratorsdegenerate,EllisonEtAl2018,RowleyMarsden2002}.
        By contrast, the ''first-discretize-then-restrict'' approach is numerically unstable, as seen in \cref{fig:VarIntegrationSSE}.
        This shows in particular that the two variational approaches do not coincide and that the first one is preferable; see Appendix \ref{sec:AppendixProofCommutativeDiagram} for further details.

        In this section, we introduced a more general approach to numerical solutions over discretized time, circumventing the need to formulate equations of motion from a restricted variational approach.
        Beyond our concrete study of entanglement in this work, the methodology can be applied to more general notions of quantumness.
        For example, all examples of classical and nonclassical systems in Ref. \cite{SW20} can be studied in this manner.
        More directly, we can even apply this approach for dynamic entanglement quantifications in terms of the Schmidt number \cite{PhysRevA.64.022306}.
        That is, it has been shown in Ref. \cite{PhysRevLett.113.260502} how the Schmidt-number quantification can be mapped to a projected separability problem to which our dynamical restriction technique can be applied straightforwardly.

\section{Conclusion}

    In this work, we utilized and developed numerical schemes for identifying dynamical entanglement through the restriction of the quantum evolution to the manifold of separable states.
    Starting from the restricted equations of motion derived from the variational formulation of separability considerations, we applied linear splitting methods to find the dynamics, demonstrating that the resulting numerical trajectories reproduce the analytical solutions in the limit of small time steps. 
    By considering a random, two-party and many-party correlation Hamiltonians, we illustrate the general applicability of the method to a broad class of physical scenarios.

    To eliminate the need for explicitly deriving the restricted equations, we introduced a variational discretization framework that directly produces integration schemes for constrained dynamics.
    Within this setting, we consider two different implementations, varying the order of the restriction and discretization steps. 
    Our numerical experiments reveal that, although both methods are formally consistent with the variational principles, the discretize–restrict approach suffers from numerical instabilities, already for the fundamental example of the exchange-interaction Hamiltonian;
    thus, the restrict-discretize order of steps is preferable.
    This observation points to a fundamental challenge in the numerical treatment of constrained dynamical systems, where the order of discretization and restriction may play a critical role in stability and accuracy.

    The results presented here establish a foundation for the numerical study of time-dependent entanglement within variational and geometric frameworks and could provide valuable computational tools for simulating the dynamics of entanglement in complex quantum systems.
    Beyond that, other restrictions apply when studying quantum effects beyond entanglement.
    The framework presented is ready to be applied in such scenarios, too.
    Thereby, the nonclassical properties of a process that drives a quantum-technological application, using a specific quantum resource, can be studied in the future.
\section*{Acknowledgments}
L.A. and J.S. acknowledge funding through the Quant\-ERA project QuCABOoSE.
\appendix
\section{Backward error analysis for Splitting schemes on SSE for the exchange interaction operator}\label{sec:AppendixBEA}

We continue in the setting of the exchange interaction (swap) operator described in the numerical experiment section \cref{sec:NumericsSwapOperator}.

Backward error analysis (BEA) seeks to find modified differential equations whose solutions coincide with the numerical solutions up to high order in the discretization parameter $\Delta t$. The structural properties of these differential equations can then be analyzed in order to better understand the behaviour of the numerical scheme. 
For the Trotter splitting scheme \eqref{eq:LieTrotterSwap} we obtain a modified differential equation as the following formal power series:
\onecolumngrid
\begin{equation}
\begin{split}\label{eq:BEATrottermodDiff}
    |\dot a \rangle &= \left(\left(-i -\frac 12 \Delta t - \frac 16 i \Delta t^2(|q|^2-1)\right)|b\rangle\langle b| + \frac 12 \Delta t |q|^2 \hat 1\right)|a \rangle + \mathcal{O}(\Delta t^3)\\
    |\dot b \rangle &= \left(\left(-i +\frac 12 \Delta t - \frac 16 i \Delta t^2(|q|^2-1)\right)|a\rangle\langle a| - \frac 12 \Delta t |q|^2 \hat 1\right)|b \rangle + \mathcal{O}(\Delta t^3).
    \end{split}
\end{equation}

\begin{figure}
    \centering
    \includegraphics[width=0.5\linewidth]{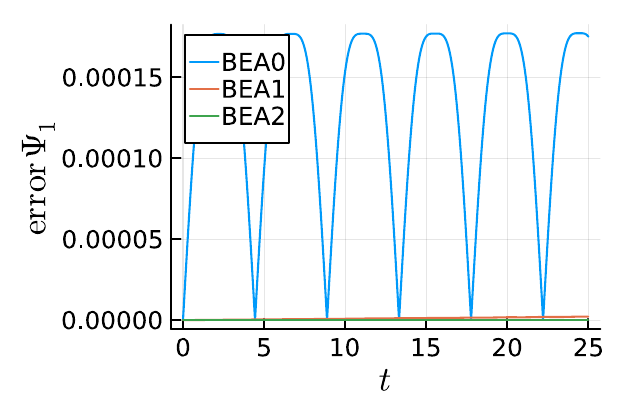}
    \includegraphics[width=0.5\linewidth]{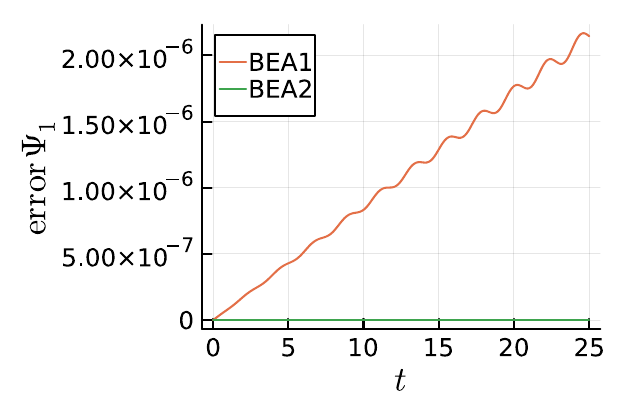}
    \includegraphics[width=0.5\linewidth]{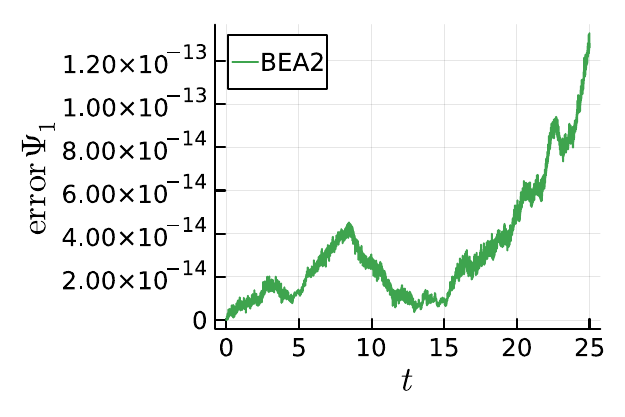}
    \caption{The difference between the solutions to truncations of \eqref{eq:BEATrottermodDiff} to the Lie-Trotter solutions are plotted for $\Delta t = 0.001$. As the plots of the three truncations are on different scales, the first plot shows all lines, the second one the last two, and the last plot only the last line.}
    \label{fig:BEATrotter}
\end{figure}

These modified equations seek to describe the behaviour of the Lie-Trotter splitting scheme. See \cref{fig:BEATrotter} for plots of solutions to truncations of \eqref{eq:BEATrottermodDiff}. The plots show that higher order truncations resemble the Lie-Trotter solutions more closely than lower order truncations. Solutions to truncations of \eqref{eq:BEATrottermodDiff} have been obtained numerically using the library {\tt DifferentialEquations.jl} \cite{rackauckas2017differentialequations} with a low error tolerance.

For any normalized initial value with transition amplitude $q$, the modified differential equation \eqref{eq:BEATrottermodDiff} coincides up to higher order terms with the solutions of the separable Schrödinger equations for the modified Hamiltonian
\[
\hat{H}^{\mathrm{mod}}_q = \left( 1 -\frac 12 i \Delta t + \frac 16 \Delta t^2 (|q|^2-1)\right)\hat H + \frac 12 i \Delta t |q|^2 \hat 1 + \mathcal{O}(\Delta t^3).
\]
(The transition amplitude $q$ is treated as a parameter when deriving the SSE for $\hat{H}^{\mathrm{mod}}_q$.)
In other words, on any levelset $q=\mathrm{const}$ the numerical solutions behave like the exact solutions to the separable Schrödinger equations for a modified Hamiltonian up to higher order terms in the discretization parameter \footnote{We refrain from discussing convergence properties of the formal power series considered in this section. In the setting of backward error analysis, convergence properties and optimal truncation strategies are discussed in \cite{HLW2013}.}. 

\begin{figure}
    \centering
    \includegraphics[width=0.5\linewidth]{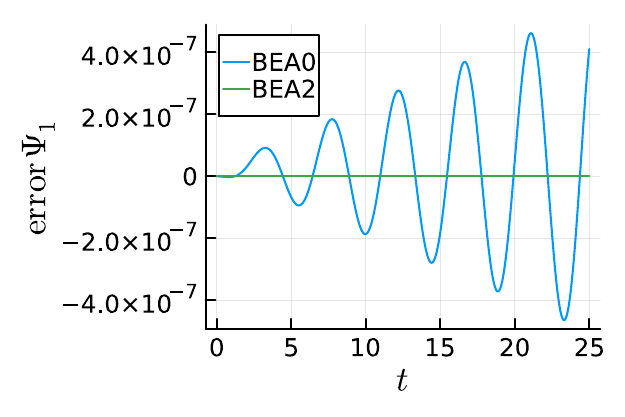}
    \includegraphics[width=0.5\linewidth]{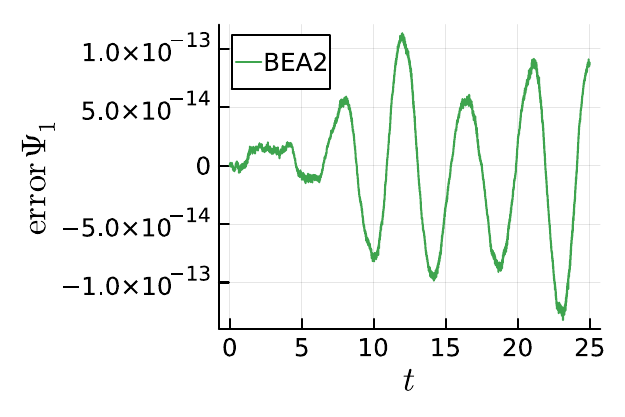}
    \caption{The difference between the solutions to truncations of \eqref{eq:BEAStrangmodDiff} to the Strang splitting solutions are plotted. As the plots of the order 0 and order 2 truncations are on different scales, the first plot shows all lines and the second one only the last one. The second order truncation resembles the Lie-Trotter solutions more closely than lower order truncations.}
    \label{fig:BEAStrang}
\end{figure}

Similarly, backward error analysis can be applied when the second-order Strang-Splitting method is applied to the SSE of the swap operator. We obtain the following modified differential equations as a formal power series in $\Delta t$:
\onecolumngrid
\begin{equation}\label{eq:BEAStrangmodDiff}
    \begin{split}
        |\dot a \rangle
        &= -i \left( \left( 1-\frac{\Delta t^2}{24}(1+2|q|^2) \right) |b \rangle \langle b| +\frac 18 \Delta t^2 |q|^2 \hat 1 \right) |a \rangle + \mathcal{O}(\Delta t^4)\\
        |\dot b \rangle
        &= -i \left( \left( 1-\frac{\Delta t^2}{24}(1-4|q|^2) \right) |a \rangle \langle a| -\frac 18 \Delta t^2 |q|^2 \hat 1 \right) |b \rangle + \mathcal{O}(\Delta t^4).
    \end{split}
\end{equation}
Notice that no terms linear in $\Delta t$ occur as the method is of second order.
\Cref{fig:BEAStrang} compares solutions of truncations of \eqref{eq:BEAStrangmodDiff} with the iterates of a Strang-Splitting integration applied to the SSE for the swap operator. 
The plots confirm that second order truncations of \eqref{eq:BEAStrangmodDiff} describe the behaviour of the Strang-Splitting method accurately on the considered time interval.

All numerical experiments can be found in \cite{GithubProject}.

\section{Discretization and separability restrictions commute at most to first order}\label{sec:AppendixProofCommutativeDiagram}

For simplicity, we consider the bipartite case as in \cref{fig:DiagramVarIntegrator}.
Consider the exact discretization in \cref{fig:DiagramVarIntegrator}, i.e.\ the discretization procedure $\Delta$ assigns the exact discrete Lagrangian. 
We obtain the two actions
\begin{align*}
	S_\Delta^{\mathrm{sep}1}(a^{(i)},b^{(i)})_i
	&= \sum_{i=0}^{N-1} \left.\int_{t_i}^{t_{i+1}} 
	L(a\otimes b, \dot a \otimes b + a \otimes \dot b) dt\right|\begin{array}{l}
		\mathrm{sep.~EL}\\
		a(t_i)=a^{(i)}\\
		b(t_i)=b^{(i)}
	\end{array}
	\\
	S_\Delta^{\mathrm{sep}2}(a^{(i)},b^{(i)})_i
	&= \sum_{i=0}^{N-1} \left.\int_{t_i}^{t_{i+1}} 
	L(\Psi,\dot \Psi) dt \right|\begin{array}{l}
		\mathrm{EL}\\
		\Psi(t_i)=a^{(i)} \otimes b^{(i)}
	\end{array}
\end{align*}
Again, we hide the complex conjugate input argument of the Lagrangian to shorten expressions.
In the $i$th summand of $S_\Delta^{\mathrm{sep}1}$, $(a,b)$ are given as solutions to separable Euler--Lagrange equations with initial condition $a(t_i)=a^{(i)}$, $b(t_i)=b^{(i)}$ for Lagrangians $L$ that define Euler--Lagrange equations of first order (as the Lagrangian of Schrödingers equation). When the Euler--Lagrange equations are of second order, then the two point boundary value problem $a(t_i)=a^{(i)}$, $b(t_i)=b^{(i)}$, $a(t_{i+1})=a^{(i+1)}$, $b(t_{i+1})=b^{(i+1)}$ is solved. This is solvable when the Lagrangian is regular \footnote{\label{foot:regularLagrangian}Denote the input variables of the Lagrangian $L$ by $q$, $\dot q$, i.e., $L(q,\dot q)$. The Lagrangian $L$ is regular, if the matrix $\frac{\partial^2 L}{\partial \dot q \partial \dot q}$ is of full rank everywhere.}
and $\Delta t >0$ is sufficiently small \cite{MarsdenWest2001}.

In contrast, in the $i$th summand of $S_\Delta^{\mathrm{sep}2}$, the unrestricted Euler--Lagrange equation is solved to obtain $\Psi$. As initial condition, $\Psi(t_i)=a^{(i)} \otimes b^{(i)}$ is used. If $L$ defines second-order differential equations, then a two-point boundary value problem with $\Psi(t_i)=a^{(i)} \otimes b^{(i)}$, $\Psi(t_{i+1})=a^{(i+1)} \otimes b^{(i+1)}$ is solved.

The action $S_\Delta^{\mathrm{sep}2}$ appears artificial as it solves the unrestricted Euler--Lagrange equation in each step. The second action is only pinned to separable states at the discrete time-points $t_i$.

The summands of $S_\Delta^{\mathrm{sep}1}$ constitute evaluations of the exact discrete Lagrangian to the Lagrangian $ L^{\mathrm{sep}}((a,b),(\bar{a},\bar{b}),(\dot a,\dot b),(\dot {\bar a},\dot {\bar b})) = L(a\otimes b,\bar a\otimes \bar b, \dot a \otimes b + a \otimes \dot b,\dot {\bar a} \otimes \bar b + \bar a \otimes \dot {\bar b})$.
In contrast, the summands of $S_\Delta^{\mathrm{sep}2}$ constitute evaluations of the exact discrete Lagrangian to $L^{\mathrm{full}}(\Psi, \bar \Psi,\dot \Psi,\dot {\bar{\Psi}})=L(\Psi, \bar \Psi,\dot \Psi,\dot {\bar{\Psi}})$.

We now consider a general variable $q$, into which we will later substitute $q=((a,b),(\bar a, \bar b))$ or $q=(\Psi, \bar{\Psi})$.
The Taylor-series expansion around $\Delta t=0$ of a general exact discrete Lagrangian $\ell_\Delta^{\mathrm{ex}}(q^{(i)},q^{(i+1)}) = \int_{t_i}^{t_{i}+\Delta t} \ell(q(t),\dot q(t)) dt$
for a variational principle to a Lagrangian $\ell(q,\dot q)$ reads

\begin{equation}
	\begin{split}
\ell_\Delta^{\mathrm{ex}}(q^{(i)},q^{(i+1)}) 
&= \int\limits^{t_i+\Delta t}_{t_i} \underbrace{\ell(q(t), \dot{q}(t))}_{\text{Taylor around }t=t_i} dt \\
		&=  \int\limits^{t_i+\Delta t}_{t_i} \left( \ell (q^{(i)}, \dot{q}^{(i)}) + t \left( \frac{\partial \ell}{\partial q} 
		(q^{(i)}, \dot{q}^{(i)}) \cdot \dot{q}^{(i)} + \frac{\partial \ell}{\partial \dot{q}} (q^{(i)}, \dot{q}^{(i)}) \cdot 
		\ddot{q}(t_i) \right) + {\cal O} ((t-t_i)^2) \right) dt \\
		&= \Delta t \ell(q^{(i)}, \dot{q}^{(i)}) + \frac{1}{2} \Delta t^2 
		\left( \frac{\partial \ell}{\partial q} (q^{(i)}, \dot{q}^{(i)}) \cdot \dot{q}^{(i)} + 
		\frac{\partial \ell}{\partial \dot{q}} (q^{(i)}, \dot{q}^{(i)}) \cdot \ddot{q}(t^{(i)}) \right) + {\cal O} (\Delta t^3).
	\end{split}
\end{equation}

When the Euler--Lagrange equations to $\ell$ are first-order differential equations, the expression $\dot{q}^{(i)}$ can be substituted by an expression that can be found by solving the Euler--Lagrange equations for $\dot{q}^{(i)}$. Derivatives of this expression can be used to obtain expressions for higher derivatives $\ddot{q}^{(i)}$ as well. 
When the Euler--Lagrange equations to $\ell$ are second-order differential equations, the expression $\ddot{q}^{(i)}$ can be substituted by an expression that can be found by solving the Euler--Lagrange equations for $\ddot{q}^{(i)}$.

Let us apply this formula to the Lagrangians $L^{\mathrm{sep}}$ and $L^{\mathrm{full}}$, where $q=((a,b),(\bar a, \bar b))$ or $q=(\Psi, \bar{\Psi})$, respectively.

When $L$ yields Euler--Lagrange equations that are first-order differential equations in time (as in the Schrödinger case), the Taylor series expansions of $L^{\mathrm{sep}}$ and $L^{\mathrm{full}}$ do not coincide to first order in $\Delta t$ in general. Indeed, the condition to coincide is that a separable state $\Psi(t_i)=a^{(i)}\otimes b^{(i)}$ is moved by $L^{\mathrm{full}}$ and by $L^{\mathrm{sep}}$ in such a way that their first derivatives at $t=t_i$ coincide. In formulas, $\dot \Psi (t_i) = \dot a(t_i) \otimes b(t_i) + a(t_i) \otimes \dot b(t_i)$, where $\Psi$ solves the unrestricted Euler--Lagrange equations with initial value $\Psi(t_i)=a^{(i)}\otimes b^{(i)}$ and corresponding initial values for the complex conjugate variables and $(a,b)$ solve the Euler--Lagrange equations to $L^{\mathrm{sep}}$ with initial value $a(t_i)=a^{(i)}$, $b(t_i)= b^{(i)}$ and corresponding initial values for the complex conjugate variables.


When $L$ yields second-order Euler--Lagrange equations (for instance, when $L$ describes a mechanical system),
the Taylor series expansions of $L^{\mathrm{sep}}$ and $L^{\mathrm{full}}$ coincide to first order in $\Delta t$ but not to second order, unless $\ddot \Psi (t_i)=\ddot a \otimes b + 2 \dot a \otimes \dot b + a \otimes \ddot b$,
where $\Psi$ solves the unrestricted Euler--Lagrange equations with initial value $\Psi(t_i)=a^{(i)}\otimes b^{(i)}$ and where $(a,b)$ solve the Euler--Lagrange equations to $L^{\mathrm{sep}}$ with initial value $\Psi(t_i)=a^{(i)}\otimes b^{(i)}$.
In other words, motions will only coincide to first order in $\Delta t$ in general.

Any variational discretization $L_\Delta$ of order $\mathcal O(\Delta t^r)$ is equivalent to a discrete Lagrangian that approximates the exact discrete Lagrangian to order $\mathcal O(\Delta t^r)$ \cite{MarsdenWest2001}.
This allows us to summarize the discussion as follows.
Consider Lagrangians in variables and their first derivatives (first-order Lagrangians).
For variational principles for degenerate Lagrangians (defining Euler--Lagrange equations that are first-order differential equations -- such as in the quantum evolution case) in general position, \cref{fig:DiagramVarIntegrator} does not commute for consistent (i.e.\ order at least $\mathcal{O}(\Delta t^1)$) discretizations.
For variational principles to regular Lagrangians (defining Euler--Lagrange equations of second order -- such as mechanical Lagrangians) in general position, the diagram \cref{fig:DiagramVarIntegrator} does not commute for discretizations of order at least $\mathcal{O}(\Delta t^2)$. 
\bibliographystyle{unsrturl}
\bibliography{literature}

@article{SV13,
  author       = {Sperling, J. and Vogel, W.},
  title        = {Multipartite Entanglement Witnesses},
  journal      = {Physical Review Letters},
  volume       = {111},
  number       = {11},
  pages        = {110503},
  year         = {2013},
  doi          = {10.1103/PhysRevLett.111.110503},
  url          = {https://doi.org/10.1103/PhysRevLett.111.110503}
}

@article{SW17,
  title = {Separable and Inseparable Quantum Trajectories},
  author = {Sperling, J. and Walmsley, I. A.},
  journal = {Phys. Rev. Lett.},
  volume = {119},
  issue = {17},
  pages = {170401},
  numpages = {6},
  year = {2017},
  month = {Oct},
  publisher = {American Physical Society},
  doi = {10.1103/PhysRevLett.119.170401},
  url = {https://link.aps.org/doi/10.1103/PhysRevLett.119.170401}
}

@article{SW20,
  author       = {Sperling, J. and Walmsley, I.\,A.},
  title        = {Classical evolution in quantum systems},
  journal      = {Physica Scripta},
  volume       = {95},
  number       = {6},
  pages        = {065101},
  year         = {2020},
  doi          = {10.1088/1402-4896/ab833b},
  url          = {https://doi.org/10.1088/1402-4896/ab833b}
}

@Article{Ceruti2024,
author="Ceruti, Gianluca
and Einkemmer, Lukas
and Kusch, Jonas
and Lubich, Christian",
title="A robust second-order low-rank BUG integrator based on the midpoint rule",
journal="BIT Numerical Mathematics",
year="2024",
month="Jul",
day="13",
volume="64",
number="3",
pages="30",
issn="1572-9125",
doi="10.1007/s10543-024-01032-x",
url="https://doi.org/10.1007/s10543-024-01032-x"
}

@article{Blanes_Casas_Murua_2024, title={Splitting methods for differential equations}, volume={33}, DOI={10.1017/S0962492923000077}, journal={Acta Numerica}, author={Blanes, Sergio and Casas, Fernando and Murua, Ander}, year={2024}, pages={1–161}
}

@book{DAmbrosio2023, title={Numerical Approximation of Ordinary Differential Problems: From Deterministic to Stochastic Numerical Methods}, ISBN={9783031313431}, ISSN={2532-3318}, url={http://dx.doi.org/10.1007/978-3-031-31343-1}, DOI={10.1007/978-3-031-31343-1}, journal={UNITEXT}, publisher={Springer Nature Switzerland}, author={D’Ambrosio, Raffaele}, year={2023} 
}

@book{LeimkuhlerReich2005, place={Cambridge}, series={Cambridge Monographs on Applied and Computational Mathematics}, title={Simulating Hamiltonian Dynamics}, publisher={Cambridge University Press}, author={Leimkuhler, Benedict and Reich, Sebastian}, year={2005}, collection={Cambridge Monographs on Applied and Computational Mathematics}
}

@article{McLachlan2002, title={Splitting methods}, volume={11}, DOI={10.1017/S0962492902000053}, journal={Acta Numerica}, author={McLachlan, Robert I. and Quispel, G. Reinout W.}, year={2002}, pages={341–434}
}

@book{HLW2013,
	Author = {Hairer, Ernst and Lubich, Christian and Wanner, Gerhard},
	Doi = {10.1007/3-540-30666-8},
	Isbn = {9783662050187},
	Publisher = {Springer Berlin Heidelberg},
	Series = {Springer Series in Computational Mathematics},
	Title = {Geometric Numerical Integration: Structure-Preserving Algorithms for Ordinary Differential Equations},
	Year = {2013}
}

@article{MarsdenWest2001,
Author = {Marsden, Jerrold E. and West, Matthew},
Doi = {10.1017/S096249290100006X},
Journal = {Acta Numerica},
Pages = {357--514},
Publisher = {Cambridge University Press},
Title = {Discrete mechanics and variational integrators},
Volume = {10},
Year = {2001}
}

@misc{kraus2017projectedvariationalintegratorsdegenerate,
title={Projected Variational Integrators for Degenerate {L}agrangian Systems}, 
author={Michael Kraus},
year={2017},
eprint={1708.07356},
archivePrefix={arXiv},
primaryClass={math.NA},
url={https://arxiv.org/abs/1708.07356}, 
doi={10.48550/arXiv.1708.07356}
}

@Article{OberBloebaum2015,
author="Ober-Bl{\"o}baum, Sina
and Saake, Nils",
title="Construction and analysis of higher order {G}alerkin variational integrators",
journal="Advances in Computational Mathematics",
year="2015",
month="Dec",
day="01",
volume="41",
number="6",
pages="955--986",
issn="1572-9044",
doi="10.1007/s10444-014-9394-8",
url="https://doi.org/10.1007/s10444-014-9394-8"
}

@book{Lubich2008,
title={From Quantum to Classical Molecular Dynamics: Reduced Models and Numerical Analysis}, 
ISBN={978-3-03719-567-3},
DOI={10.4171/067}, 
publisher={European Mathematical Society}, 
author={Lubich, Christian}, 
year={2008} 
}

@INPROCEEDINGS{RowleyMarsden2002,
author={Rowley, Clarence W. and Marsden, Jerrold E.},
booktitle={Proceedings of the 41st IEEE Conference on Decision and Control, 2002.}, 
title={Variational integrators for degenerate {Lagrangians}, with application to point vortices}, 
year={2002},
volume={2},
number={},
pages={1521-1527},
doi={10.1109/CDC.2002.1184735}
}

@misc{GithubProject,
author={},
title={GitHub Repository Christian-Offen/SeparableSchroedinger},
    note = {Source code for the numerical experiments and backward error analysis.},
url={https://github.com/Christian-Offen/SeparableSchroedinger},
doi={10.5281/zenodo.18483123}
}

@article{rackauckas2017differentialequations,
  title={Differential{E}quations.jl--a performant and feature-rich ecosystem for solving differential equations in {J}ulia},
  author={Rackauckas, Christopher and Nie, Qing},
  journal={Journal of Open Research Software},
  volume={5},
  number={1},
  year={2017},
  publisher={Ubiquity Press}
}

@article{ISERLES202429,
title = {An elementary approach to splittings of unbounded operators},
journal = {Computers {\&} Mathematics with Applications},
volume = {176},
pages = {29-34},
year = {2024},
issn = {0898-1221},
doi = {https://doi.org/10.1016/j.camwa.2024.08.031},
url = {https://www.sciencedirect.com/science/article/pii/S0898122124003985},
author = {Arieh Iserles and Karolina Kropielnicka}
}

@book{GR96,
  author    = {W. Greiner and J. Reinhardt},
  title     = {Field Quantization},
  publisher = {Springer},
  year      = {1996},
  doi       = {10.1007/978-3-642-61485-9},
  url       = {https://doi.org/10.1007/978-3-642-61485-9}
}

@article{YS24,
   title={{Entanglement-assisted quantum speedup: Beating local quantum speed limits}},
   volume={110},
   ISSN={2469-9934},
   url={http://dx.doi.org/10.1103/PhysRevA.110.012424},
   DOI={10.1103/physreva.110.012424},
   number={1},
   journal={Physical Review A},
   publisher={American Physical Society (APS)},
   author={Yasmin, F. and Sperling, J.},
   year={2024},
   month=jul 
}

@article{I07,
  author = {L. M. Ioannou},
  title = {Computational Complexity of the Quantum Separability Problem},
  journal = {Quantum Information I\& Computation},
  volume = {7},
  number = {4},
  pages = {335--370},
  year = {2007},
  doi = {10.26421/QIC7.4-5},
  url = {https://doi.org/10.26421/QIC7.4-5}
}

@article{GVS18,
  title = {Numerical Construction of Multipartite Entanglement Witnesses},
  author = {Gerke, S. and Vogel, W. and Sperling, J.},
  journal = {Phys. Rev. X},
  volume = {8},
  issue = {3},
  pages = {031047},
  numpages = {18},
  year = {2018},
  month = {Aug},
  publisher = {American Physical Society},
  doi = {10.1103/PhysRevX.8.031047},
  url = {https://link.aps.org/doi/10.1103/PhysRevX.8.031047}
}

@article{P96,
  title={Separability criterion for density matrices},
  author={Peres, Asher},
  journal={Physical Review Letters},
  volume={77},
  number={8},
  pages={1413--1415},
  year={1996},
  publisher={APS}
}

@article{H09,
  title={Quantum entanglement},
  author={Horodecki, Ryszard and Horodecki, Pawe{\l} and Horodecki, Micha{\l} and Horodecki, Karol},
  journal={Reviews of Modern Physics},
  volume={81},
  number={2},
  pages={865--942},
  year={2009},
  publisher={APS}
}

@article{GT09,
  title={Entanglement detection},
  author={G{\"u}hne, Otfried and T{\'o}th, G{\'e}za},
  journal={Physics Reports},
  volume={474},
  number={1-6},
  pages={1--75},
  year={2009},
  publisher={Elsevier}
}

@article{PhysRevLett.99.120501,
  title = {Nonclassical Interference and Entanglement Generation Using a Photonic Crystal Fiber Pair Photon Source},
  author = {Fulconis, J\'er\'emie and Alibart, Olivier and O'Brien, Jeremy L. and Wadsworth, William J. and Rarity, John G.},
  journal = {Phys. Rev. Lett.},
  volume = {99},
  issue = {12},
  pages = {120501},
  numpages = {4},
  year = {2007},
  month = {Sep},
  publisher = {American Physical Society},
  doi = {10.1103/PhysRevLett.99.120501},
  url = {https://link.aps.org/doi/10.1103/PhysRevLett.99.120501}
}

@article{PhysRevLett.101.170502,
  title = {Entanglement Evolution in Finite Dimensions},
  author = {Tiersch, Markus and de Melo, Fernando and Buchleitner, Andreas},
  journal = {Phys. Rev. Lett.},
  volume = {101},
  issue = {17},
  pages = {170502},
  numpages = {4},
  year = {2008},
  month = {Oct},
  publisher = {American Physical Society},
  doi = {10.1103/PhysRevLett.101.170502},
  url = {https://link.aps.org/doi/10.1103/PhysRevLett.101.170502}
}

@article{doi:10.1142/S1230161209000153,
author = {Isar, Aurelian},
title = {Entanglement Generation and Evolution in Open Quantum Systems},
journal = {Open Systems \& Information Dynamics},
volume = {16},
number = {02n03},
pages = {205-219},
year = {2009},
doi = {10.1142/S1230161209000153},

URL = {https://doi.org/10.1142/S1230161209000153},
eprint = {https://doi.org/10.1142/S1230161209000153}
}

@article{PhysRevA.65.012101,
  title = {Dynamics of quantum entanglement},
  author = {\ifmmode \dot{Z}\else \.{Z}\fi{}yczkowski, Karol and Horodecki, Pawe\l{} and Horodecki, Micha\l{} and Horodecki, Ryszard},
  journal = {Phys. Rev. A},
  volume = {65},
  issue = {1},
  pages = {012101},
  numpages = {9},
  year = {2001},
  month = {Dec},
  publisher = {American Physical Society},
  doi = {10.1103/PhysRevA.65.012101},
  url = {https://link.aps.org/doi/10.1103/PhysRevA.65.012101}
}

@article{CLMOW14,
  title = {Everything You Always Wanted to Know About LOCC (But Were Afraid to Ask)},
  author = {Chitambar, Eric and Leung, Debbie and Mančinska, Laura and Ozols, Maris and Winter, Andreas},
  journal = {Commun. Math. Phys.},
  volume = {328},
  pages = {303–326},
  year = {2014},
  month = {Mar},
  doi = {10.1007/s00220-014-1953-9},
  url = {https://doi.org/10.1007/s00220-014-1953-9}
}

@article{PhysRevA.86.050302,
  title = {Multipartite entanglement evolution under separable operations},
  author = {Gheorghiu, Vlad and Gour, Gilad},
  journal = {Phys. Rev. A},
  volume = {86},
  issue = {5},
  pages = {050302},
  numpages = {5},
  year = {2012},
  month = {Nov},
  publisher = {American Physical Society},
  doi = {10.1103/PhysRevA.86.050302},
  url = {https://link.aps.org/doi/10.1103/PhysRevA.86.050302}
}

@article{PhysRevA.88.042335,
  title = {Quantum channel detection},
  author = {Macchiavello, C. and Rossi, M.},
  journal = {Phys. Rev. A},
  volume = {88},
  issue = {4},
  pages = {042335},
  numpages = {6},
  year = {2013},
  month = {Oct},
  publisher = {American Physical Society},
  doi = {10.1103/PhysRevA.88.042335},
  url = {https://link.aps.org/doi/10.1103/PhysRevA.88.042335}
}

@article{PhysRevResearch.4.013200,
  title = {Quantifying the entanglement of quantum channel},
  author = {Zhou, Huaqi and Gao, Ting and Yan, Fengli},
  journal = {Phys. Rev. Res.},
  volume = {4},
  issue = {1},
  pages = {013200},
  numpages = {10},
  year = {2022},
  month = {Mar},
  publisher = {American Physical Society},
  doi = {10.1103/PhysRevResearch.4.013200},
  url = {https://link.aps.org/doi/10.1103/PhysRevResearch.4.013200}
}

@misc{pinske2025entanglingpowernonentanglingchannels,
      title={The entangling power of non-entangling channels}, 
      author={Julien Pinske and Jan Sperling and Klaus Mølmer},
      year={2025},
      eprint={2512.14819},
      archivePrefix={arXiv},
      primaryClass={quant-ph},
      url={https://arxiv.org/abs/2512.14819}, 
}

@article{PA24,
  title = {Separability {Lindblad} equation for dynamical open-system entanglement},
  author = {Pinske, Julien and Ares, Laura and Hinrichs, Benjamin and Kolb, Martin and Sperling, Jan},
  journal = {Phys. Rev. A},
  volume = {113},
  issue = {1},
  pages = {L010403},
  numpages = {6},
  year = {2026},
  month = {Jan},
  publisher = {American Physical Society},
  doi = {10.1103/kd3b-bfxq},
  url = {https://link.aps.org/doi/10.1103/kd3b-bfxq}
}

@article{AP24,
  title = {Restricted {Monte Carlo} wave-function method and {Lindblad} equation for identifying entangling open-quantum-system dynamics},
  author = {Ares, Laura and Pinske, Julien and Hinrichs, Benjamin and Kolb, Martin and Sperling, Jan},
  journal = {Phys. Rev. A},
  volume = {113},
  issue = {1},
  pages = {012220},
  numpages = {13},
  year = {2026},
  month = {Jan},
  publisher = {American Physical Society},
  doi = {10.1103/hcj7-8zlg},
  url = {https://link.aps.org/doi/10.1103/hcj7-8zlg}
}

@article{PhysRevA.64.022306,
  title = {Schmidt measure as a tool for quantifying multiparticle entanglement},
  author = {Eisert, Jens and Briegel, Hans J.},
  journal = {Phys. Rev. A},
  volume = {64},
  issue = {2},
  pages = {022306},
  numpages = {4},
  year = {2001},
  month = {Jul},
  publisher = {American Physical Society},
  doi = {10.1103/PhysRevA.64.022306},
  url = {https://link.aps.org/doi/10.1103/PhysRevA.64.022306}
}

@article{PhysRevLett.113.260502,
  title = {Structural Quantification of Entanglement},
  author = {Shahandeh, F. and Sperling, J. and Vogel, W.},
  journal = {Phys. Rev. Lett.},
  volume = {113},
  issue = {26},
  pages = {260502},
  numpages = {5},
  year = {2014},
  month = {Dec},
  publisher = {American Physical Society},
  doi = {10.1103/PhysRevLett.113.260502},
  url = {https://link.aps.org/doi/10.1103/PhysRevLett.113.260502}
}

@article{EllisonEtAl2018,
    author = {Ellison, C. L. and Finn, J. M. and Burby, J. W. and Kraus, M. and Qin, H. and Tang, W. M.},
    title = {Degenerate variational integrators for magnetic field line flow and guiding center trajectories},
    journal = {Physics of Plasmas},
    volume = {25},
    number = {5},
    pages = {052502},
    year = {2018},
    month = {05},
    issn = {1070-664X},
    doi = {10.1063/1.5022277},
    url = {https://doi.org/10.1063/1.5022277},
    eprint = {https://pubs.aip.org/aip/pop/article-pdf/doi/10.1063/1.5022277/14689237/052502_1_online.pdf},
}
\end{document}